\documentclass[amsmath,amssymb,floatfix,prl]{revtex4}

\usepackage[pdftex]{graphicx}
\usepackage{amssymb,amsfonts,amsmath}
\usepackage{graphicx,amsmath}
\usepackage{float}

\begin{document}

\title{Floquet theory of microwave absorption by an impurity in two dimensional electron gas}

\author{Alexei D. Chepelianskii}
\affiliation{LPS, Universit\'e Paris-Sud, CNRS, UMR 8502, Orsay F-91405, France.}
\author{Dima L. Shepelyansky}
\affiliation{Laboratoire de Physique Th\'eorique du CNRS, IRSAMC, 
Universit\'e de Toulouse, UPS, Toulouse 31062, France}

\date{\today}

\begin{abstract}
We investigate the dynamics of a two-dimensional electron gas (2DEG) under circular 
polarized microwave radiation in presence of dilute localized impurities.
Inspired by recent developments on Floquet topological insulators 
we obtain the Floquet wavefunctions of this system
which allow us to predict the microwave absorption and charge density responses 
of the electron gas, we demonstrate how these properties 
can be understood from the underlying semiclassical dynamics even for impurities with a size of around a magnetic length. 
The charge density response takes the form of a rotating charge density vortex around the impurity 
that can lead to a significant renormalization of the external microwave field which
becomes strongly inhomogeneous on the scale of a cyclotron radius around the impurity. 
We show that this in-homogeneity can suppress the circular polarization 
dependence which is theoretically expected for MIRO but 
which was not observed in MIRO experiments on semiconducting 2DEGs. 
Our explanation, for this so far unexplained polarization independence, 
has close similarities with the Azbel'-Kaner effect in metals where 
the interaction length between the microwave field and conduction electrons 
is much smaller than the cyclotron radius due to skin effect generating harmonics of the cyclotron resonance.
\end{abstract}

\maketitle

\subsection{I Introduction.} 
The topological properties of condensed matter systems have been the focus of intense theoretical and 
experimental research starting from the discovery of the quantum Hall effect \cite{Hall1,Hall2}. 
Recently it has been suggested 
that some systems can develop new topological properties under external periodic driving, leading to 
the concept of Floquet topological insulators  \cite{Demler,Galitski,Gawędzki} and to their realization 
in photonic systems \cite{Segev1,Wang,Segev2}. 
In parallel an active research on the effect of microwave irradiation of ultra-high mobility two-dimensional systems 
revealed a striking microwave induced resistance oscillations (MIRO) 
at weak non quantizing magnetic fields \cite{Zudov1}. 
These oscillations are characterized by a periodic dependence on the ratio between 
the microwave and cyclotron frequencies.
 At sufficiently high microwave power the oscillations grow in amplitude leading to the formation 
of zero-resistance states \cite{Mani,Zudov2}. This effect has now been observed 
in several systems \cite{Denis1,Bykov,ZudovGe,ZnO,Denis2} 
and it has been shown that the zero-resistance regime can lead 
to the onset of new thermodynamic properties 
like an incompressible behavior usually associated with the quantum Hall effect \cite{Natcom}. 
The theoretical understanding of MIRO 
has attracted a considerable attention stimulating several approaches. Semiclassical and 
quantum kinetic equation formalisms have been used to derive 
an analytic description for the magnetic field dependence of 
microwave induced oscillations due to interactions with residual impurities 
in the two-dimensional electron gas in a 
Born approximation limit where re-collision or memory effects 
were treated perturbatively \cite{Theory1,Girvin,TheoryIvan1,TheoryIvan2,Zudovrmp}. 
Classical dynamics has been used to analyze memory effects in the non-perturbative 
limit and to derive in a transparent manner 
the results from kinetic equation calculations \cite{Zhirov,Dyakonov}. 
These theories provide a good description of the experimental magnetic field dependencies. 
However  they are intrinsically linked to the cyclotron-resonance and share its strong dependence  
on the polarization of the radiation. This simulated careful polarisation dependent experiments \cite{Smet,Mani1,Mani2,Kvon,Ganichev},
in particular it was shown that contrarily to cyclotron resonance MIRO is not sensitive to the orientation of circularly polarized microwave irradiation \cite{Smet,Kvon}. This disagreement stimulated the development of extrinsic theories where the influence of edges and 
contacts was considered as the main mechanism behind MIRO \cite{Chepelianskii,Mikhailov}. 
But recent experiments with a spatially resolved 
THz excitation \cite{Kvon,Ganichev} have confirmed the bulk origin of MIRO leaving the circularly 
polarized radiation dependence unexplained. 

Here we address this issue by developing a Floquet description of the interaction between a local impurity 
and a two-dimensional electron gas (2DEG) under circularly polarized radiation. 
We show that the properties of quantum Floquet wavefunctions can be understood from classical dynamics 
even in the limit where the radius of the impurity is of the order of the magnetic length. 
We then use this theory to compute the absorbed microwave power using both quantum 
and semiclassical formalisms and show the appearance of a rotating redistribution of electron density around 
the impurity that locally enhances the external electromagnetic field. 
We argue that the inclusion of this screening correction suppresses 
the dependence on polarization chirality and discuss analogies with the cyclotron resonance oscillations 
in metals known as the Azbel'-Kaner effect \cite{Kaner,Azbel} 
(see also overview of this effect in \cite{Kittel}). 

\subsection{II Model description and Dynamics - Quantum and classical Theory}

We consider an isolated impurity with a rotation invariant potential $V_i(r)$ 
embedded in a surrounding 2DEG under a magnetic field $B$ giving rise to cyclotron motion 
with a frequency $\omega_c = q B/m$ and a cyclotron radius $R_c=v_F/\omega_c$ 
($m$ and $q$ are the carrier effective mass and charge, 
$v_F$ and $E_F=m v_F^2/2$ are the Fermi velocity and energy,
$r$ is the radial distance from the impurity center).
A circularly polarized irradiation induces an additional time-dependent potential 
$V_{ac}(t) = -q E_{ac} r \cos(\theta - \omega t)$ where $\omega$ is the microwave frequency 
and $\theta$ is the polar angle. The system Hamiltonian  is the sum of 
static and time-dependent contributions and reads:
\begin{align}
{\hat H}(t) = {\hat H}_0 + V_i(r) - q E_{ac} r \cos(\theta - \omega t)
\label{eq:ht}
\end{align}
where ${\hat H}_0$ is the charge kinetic energy in a magnetic field. This time dependent Hamiltonian 
can be analyzed using general Floquet-states theory, however as the static part of 
the Hamiltonian has rotational invariance it is more convenient to move first  to 
the rotating frame where the Hamiltonian becomes stationary. This transformation is analogous to 
the rotating frame approximation in spin-resonance or in atomic physics 
which becomes exact for a circularly polarized excitation. For this purpose we introduce 
the rotating frame angle $\theta_{\cal R} = \theta - \omega t$ and seek the solutions of 
the Schr\"odinger equation in the form $\psi = |\psi_{\cal R}(r,\theta_{\cal R})> e^{-i \epsilon t/\hbar}$. 
The wavefunction $\psi_{\cal R}$ then obeys 
a stationary Schr\"odinger equation with a modified Hamiltonian ${\hat H}_{\cal R}$:
\begin{align}
{\hat H}_{\cal R} = {\hat H}_0 + V_i(r) - q E_{ac} r \cos \theta_{\cal R} - \omega {\hat l}_z 
\label{eq:hr}
\end{align}
where ${\hat l}_z$ is the orbital momentum operator defined from the position of center of the impurity, 
by construction it is conjugated to the phase $\theta_{\cal R}$. The eigenfunctions $\psi_{n}$ of 
this Hamiltonian ${\hat H}_{\cal R}$ give the Floquet wavefunctions in the laboratory frame 
$\psi = \psi_n(r,\theta_{\cal R} = \theta - \omega t)$, they correspond to wavefunctions rotating at 
frequency $\omega$. Since the rotating frame Hamiltonian is stationary 
it is possible to analyze the properties 
of the Floquet eigenstates using the technique of 
Wigner functions and associated Husimi representations \cite{Haake} 
that allows to draw a parallel between the quantum evolution and conceptually simpler classical dynamics. 
For example, we will show that the classical description leads to a straightforward description of 
the periodic dependence of transport properties on the frequency ratio $J = \omega/\omega_c$. 
The methods of numerical solution of the Schr\"odinger equation and Newton dynamics
of the Hamiltonian (\ref{eq:hr}) are described in  Supplementary Information (SI) Sec.1, Sec.2, Sec.3, Sec.4.

\begin{figure}[h]
\includegraphics[width=0.30\columnwidth]{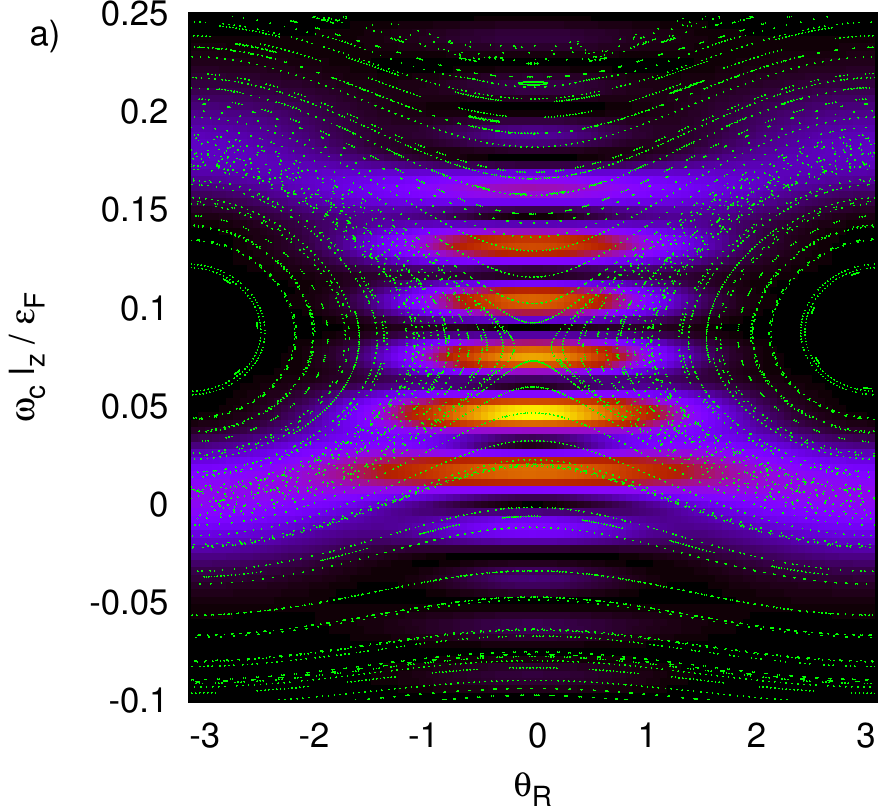}
\includegraphics[width=0.30\columnwidth]{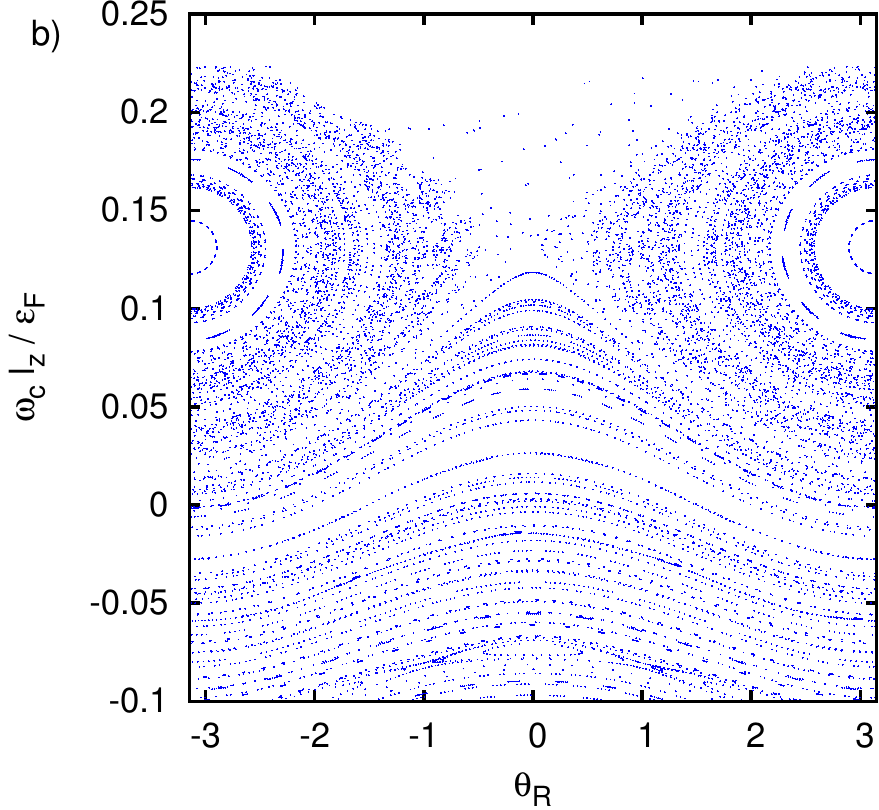}
\includegraphics[width=0.30\columnwidth]{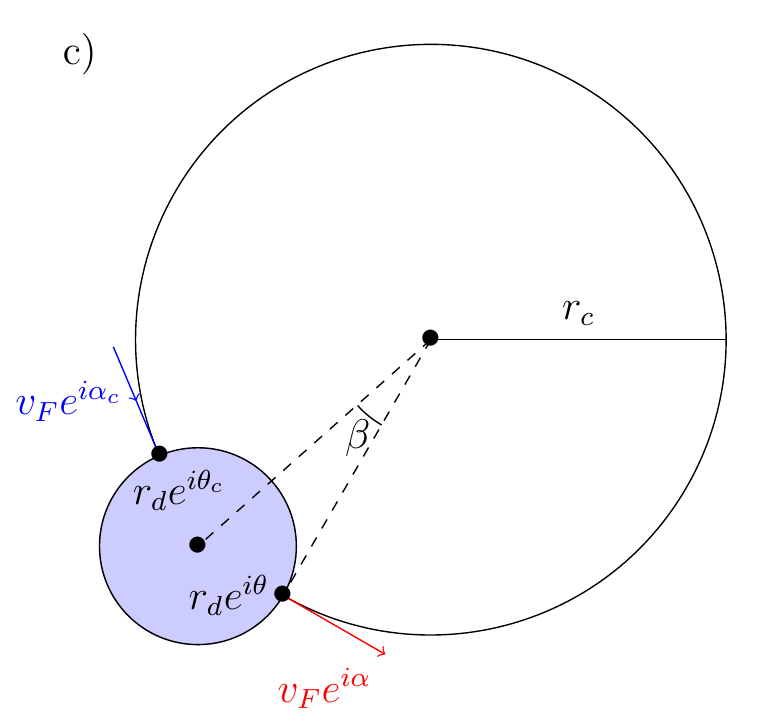}
\caption{\label{fig1}
Left panel (a): density of Husimi function at the disc radius $r=r_d$ shown by color (maximum for yellow/white;
minimum zero for violet/black), Poincar\'e section for classical Hamiltonian dynamics of (\ref{eq:hr})
is drown at the moment of collision with disc and is shown by green dots; here $J=2.7$, $r_d/\ell_B=1$, $q E_{ac} \ell_B/\hbar \omega_c =0.3$ at energy $\epsilon_F / \hbar \omega_c \approx 40$. 
Middle panel (b): classical Poincar\'e section of the collision map (\ref{eq:map}) at parameters of panel (a).
Right panel (c): geometry of collision. 
}
\end{figure}

Fig.~\ref{fig1}(a) shows the Husimi representation of a typical eigenstate of ${\hat H}_{\cal R}$ 
for a hard disc potential with radius $r_d / \ell_B = 1$ (we introduce the magnetic length 
$\ell_B = \sqrt{\hbar/m \omega_c}$) under microwave irradiation at $J = 2.7$ and 
microwave excitation amplitude $q E_{ac} \ell_B / \hbar \omega_c = 0.3$. It gives 
the semiclassical probability density in the phase space $l_z, \theta_R$ in the vicinity of 
the impurity at $r \approx r_d$. It is computed by numerical diagonalisation of 
the Hamiltonian ${\hat H}_{\cal R}$ and using the approach described in \cite{Husimi1,Husimi2} and SI Sec.2. 
The Husimi density highlights a resonant structure with a significant variation of orbital momentum 
$l_z$ with the conjugated phase $\theta_R$. The variation of orbital momentum is due to 
the action of the microwave field which spoils orbital momentum conservation
existing  at $E_{ac} = 0$. On top of the quantum Husimi distribution we have overlaid 
the corresponding Poincar\'e section of classical orbits 
\cite{Lichtenberg} obtained from the  classical dynamics described by $H_{\cal R}$. 
The results demonstrate that the resonant structure in the $l_z, \theta_R$ plane is accurately reproduced 
by classical dynamics even if the radius of the disc is comparable with the quantum magnetic length. 

The resonant structure in phase space revealed by the Husimi distribution and associated Poincar\'e section 
can be understood by computing the change in dynamical variables after 
each free evolution cyclotron period and subsequent collision with the impurity. 
We assume that the impurity potential $V_i(r)$ vanishes outside a characteristic typical radius 
$r_d \ll v_F/\omega_c$ and note by $\alpha$ the polar angle of the velocity in 
the laboratory frame when an electron leaves the interaction region $r \le r_d$. 
The images of typical trajectory collisions with the disc can be found in \cite{Zhirov}
(see also Fig.~\ref{fig1}(c) and Fig.S1 in SI Sec.4).
Without microwaves the change of $\alpha$ to its value ${\bar \alpha}$, 
taken after a cyclotron orbit rotation 
and a re-collision event with the impurity, can be expressed as 
\begin{align}
{\bar \alpha} = \alpha + \sigma(l_z)
\label{eq:alpha}
\end{align}
where $\sigma$ is a function depending on the impurity potential and on the orbital momentum giving 
the impact parameter with the impurity (the geometry of collision is shown in Fig.~\ref{fig1}(c) ). For a hard disc impurity ${\bar \alpha} = \alpha + 2 \chi - \pi$, 
where $\chi = \alpha - \theta$ is relative angle between the velocity and the impact position; 
$\chi$ is related to orbital momentum through $l_z = m v_F r_d \sin \chi$. A more general expression of 
$\sigma(l_z)$ valid for a step potential of fixed height $U_i$ is given in SI Sec.4.

Without microwaves the orbital momentum $l_z$ and the total energy $H_0$ are conserved. 
In presence of microwaves however $H_{\cal R}$ is the only integral of motion. 
The change of orbital momentum during the free-evolution time between two 
successive collisions can thus be expressed as:
\begin{align}
\delta l_z = \frac{\delta H_0}{\omega} = \frac{1}{\omega}  
\int_0^{2 \pi/\omega_c} q \mathbf{E}_{ac}(t) \mathbf{v}(t) dt  
= \frac{q v_F E_{ac}}{\omega (\omega - \omega_c)} 
\left[ \sin \alpha_R - \sin(\alpha_R - 2 \pi J) \right]
\label{eq:deltalz}
\end{align}
where we introduced $\alpha_{\cal R}$ the velocity polar angle in the rotating frame 
$\alpha_{\cal R}= \alpha - \omega t$, $J = \omega/\omega_c$.
During the short collision with the impurity we can neglect the microwave field so that
in this approximation the orbital momentum is conserved in the interaction region. 
Combining the two previous equations we find a symplectic map \cite{Lichtenberg,Chirikov} 
describing the evolution of the velocity polar angle $\alpha_{\cal R}$ and conjugated
orbital momentum $\l_z$ from one collision to another:
\begin{align}
\left\{
\begin{array}{cl}
{\bar l}_z &= l_z +  F \left[ \sin \alpha_R - \sin(\alpha_R - 2 \pi J) \right] \;, \\
{\bar \alpha}_{\cal R} &= \alpha_{\cal R} + \sigma({\bar l}_z) - 2 \pi J \;, \;\; 
F = q v_F E_{ac}/[\omega (\omega - \omega_c)] \; .
\end{array}
\right.
\label{eq:map}
\end{align}
Here bars mark the dynamical values of variables after one map iteration (one collision) 
and $\sigma$ is a function describing the change in the angle of the velocity after collision with 
an impurity with an impact parameter given by $l_z$. For colliding trajectories, the orbital momentum is given 
by $l_z = m v_F r_d \sin \chi$ where $\chi$ is the angle between the impact position on the impurity 
and the impact velocity (see SI Sec.4). For a hard disc potential we find $\sigma = 2 \chi - \pi$ 
whereas for a strong attractive potential $\sigma$ is $\pi$ shifted and is given by $\sigma = 2 \chi$. 
The behaviour of the map then depends on the dimensionless force 
$\epsilon =  q E_{ac}/[m r_d \omega (\omega - \omega_c)]$ 
which can be strong even for weak amplitudes of the microwave field
since $r_d$ appears in the denominator. More details on map derivation are presented in SI Sec.4.

This map gives a good description of the dynamics leading to practically the same Poincar\'e section 
as exact Hamiltonian evolution 
(see Fig.~\ref{fig1}(b)) and 
allows to understand the physical origin of the oscillatory dependence on the parameter $J = \omega/\omega_c$. 
In the frame of the free-particle moving under magnetic and microwave fields, the microwave leads 
effectively to a vibration of the impurity position, the change in orbital momentum thus depends 
on the position of the impurity at the moment of re-collisions. In the limit of a small impurity 
$r_d \ll R_c = v_F/\omega_c$ the time between recollisions will be given by the cyclotron period and 
the position of the impurity in the vibrating frame will depend periodically on $J$. 
For integer values of $J$ the position of the impurity does not change from collision 
to collision leading to a vanishing kick $\delta l_z \propto [ \sin \alpha_R - \sin(\alpha_R - 2 \pi J) ]$ 
as derived in Eq.~(\ref{eq:deltalz}).

This result may seem surprising from a quantum point of view, 
since one could expect strong effect of microwave 
at integer harmonics of the cyclotron resonance as it would to correspond to resonant absorption of photons 
between separated Landau levels. This issue can be elucidated through the computation of 
the absorbed microwave power in both quantum and semiclassical cases using the quantum Floquet master equation 
and classical kinetic equation and approximate map calculations. This calculation will also reveal 
the connection between the absorbed microwave power and an effective rotating dipole that appears due 
to the formation of a charge density vortex around the impurity in the rotating frame. This will then give some insight 
on the possible origin of the so far unexplained circular polarization dependence in MIRO experiments.

\subsection{III Master and Kinetic equations}

The Floquet eigenstates form a natural basis to write the master equation which describes excitation 
of the 2DEG by microwaves and relaxation to equilibrium \cite{Kohler,Hanggi}. As we would like to focus on 
the physical properties of the Floquet eigenstates around an impurity, 
we assume a simplified master equation 
in a relaxation time approximation (more sophisticated relaxation models have been studied in the limit 
of semiclassical quantum kinetic theory but without computing the exact Floquet states \cite{Zudovrmp}). 
The master equation for the density matrix ${\hat \rho}(t)$ in the relaxation time approximation then reads:
\begin{align}
\frac{\partial {\hat \rho}}{\partial t} = -\frac{i}{\hbar} [ {\hat H}(t), {\hat \rho} ] 
- \frac{ {\hat \rho} - {\hat \rho}_{eq} }{\tau} 
\label{eq:master1}
\end{align}

The equilibrium density matrix  ${\hat \rho}_{eq}$ is the equilibrium Fermi-Dirac distribution 
for the Hamiltonian without microwaves.
Following general Floquet theory, we write this equation in the basis of the Floquet eigenstates. 
In our problem those are the eigenfunctions $|\psi_{\cal R}^{(n)}>$ of the rotating frame Hamiltonian 
${\hat H}_{\cal R}$. We find that the matrix elements 
${\hat \rho}_{nm} = <\psi_{\cal R}^{(n)}| {\hat \rho} | \psi_{\cal R}^{(m)}>$ obey the following equation:
\begin{align}
\frac{\partial}{\partial t} \rho_{nm} + \frac{i }{\hbar} \epsilon_{nm} \rho_{nm} 
= - \frac{\rho_{nm} - \rho_{eq,nm}}{\tau}
\label{eq:master2}
\end{align}
where $\epsilon_{nm} = \epsilon_n - \epsilon_m$ is the difference of associated quasi-energies. 

The quantity $\rho_{eq,nm} = <\psi_{\cal R}^{(n)}| {\hat \rho}_{eq} | \psi_{\cal R}^{(m)}>$ 
can a-priori depend on time 
due to the time dependence of the Floquet states $\psi_{\cal R}$. 
However these states are stationary in the rotating frame 
$\theta_{\cal R} = \theta - \omega t$ and the steady state Hamiltonian  ${\hat H}_0$ 
is rotation invariant. Thus, as it can be shown by an explicit calculation, 
the matrix elements $\rho_{eq,nm}$ are actually independent of time. 

At large times $t \gg \tau$, the density matrix ${\hat \rho}$ converges 
to a stationary solution in the rotating frame basis:
\begin{align}
\rho_{nm} = \frac{ \rho_{eq,nm} }{ 1 + \frac{i}{\hbar} \epsilon_{nm} \tau  }
\label{eq:masterstat}
\end{align}
This steady state density matrix can  be used to compute several physical properties: 
microwave absorption, redistribution of charges around the impurity, induced rotating dipole 
appearing near the impurity due to the charge redistribution in the rotating frame. 

To deepen our understanding of the link between quantum and classical dynamics, 
we also solve the associated kinetic Vlasov equation for the distribution function in the classical phase space:
\begin{align}
\frac{\partial f}{\partial t} + \left\{f, H(t)\right\} = - \frac{f - f_{eq}}{\tau}
\label{eq:kinetic}
\end{align}
where we introduced the Poisson-brackets defined as
\begin{align}
\left\{f, H\right\}  =  \frac{\partial f}{\partial \mathbf{r}} \frac{\partial H}{\partial {\mathbf p}} 
 -  \frac{\partial f}{\partial \mathbf{p}} \frac{\partial H}{\partial {\mathbf r}}
\label{eq:poissonbrack}
\end{align}
As in the quantum case, the kinetic equation becomes time independent in the rotating frame. 
In the rotating frame the distribution function $f_{\cal R}$ is found numerically 
by integrating the characteristic equation 
$\frac{d f_{\cal R}}{d t} = - \frac{f_{\cal R} - f_{eq}}{\tau}$ 
along trajectories in this frame. 

The details of numerical simulations of the quantum master equation and
the classical kinetic equation are given in SI Sec.1, Sec.3, Sec.5.

\subsection{IV Microwave power absorption by electrons interacting with an impurity}

The microwave power absorbed by electrons interacting with an isolated impurity can be obtained 
by averaging the operator $\mathbf{v}\cdot q\mathbf{E}_{ac}$ in the rotating frame. 
In the quantum case, this is done by computing the trace: 
${\cal P} = {\rm Tr} ({\hat \rho} \mathbf{v}\cdot q\mathbf{E}_{ac} )$, 
for the Vlasov equation a similar procedure, described in supplementary information, is employed. 
\begin{figure}[h]
\centering
\input{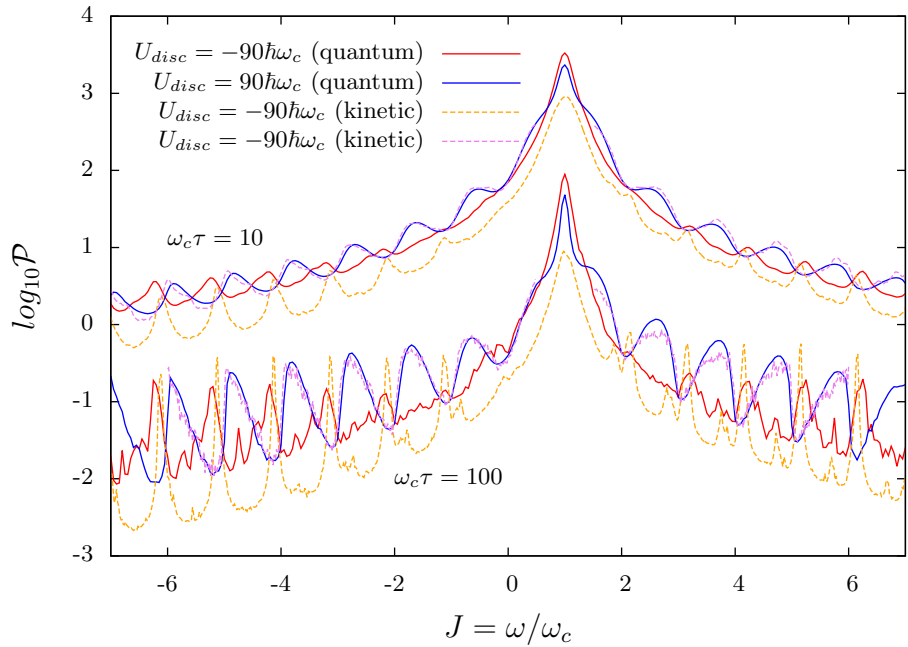}
\caption{\label{fig2}
Absorption power  ${\cal P}$ (in arbitrary units) is shown as 
a function of $J=\omega/\omega_c$ for $r_d/\ell_B=2$ at
$qE_{ac}\ell_B/\hbar \omega_c=0.1$ and $\epsilon_F/\hbar \omega_c=60$.
Data are shown for two relaxation times with $\omega_c \tau=10, 100$ (different signs of $J$ correspond to different signs of circular polarization). 
Quantum results are obtained from the master equation on the density matrix, they are compared with classical results from the kinetic equation. An excellent agreement between quantum and classical results is observed for a repulsive potential. For an attractive potential the agreement is only qualitative the kinetic equation giving the same phase of microwave absorption oscillations as the quantum calculation. 
}
\end{figure}

The dependence of the absorbed power ${\cal P}$ on $J$ is shown in Fig.~\ref{fig2} for a
strongly repulsive ($U_{disc} > E_F >0$) and attractive  ($U_{disc} < -E_F <0$) impurity potential, different signs of $J$ correspond to different signs of circular polarization. The absorbed power shows a resonance at $J = 1$ which corresponds to cyclotron resonance (which appears only for the active polarisation $J > 0$), ${\cal P}$ then decays as expected from an usual Drude model but presents additional oscillations with a period $\Delta J = 1$ which become more pronounced as $\omega_c \tau$ grows.  These oscillations appear due to the periodic structure in $J$ in Eq.~(\ref{eq:map}) and correspond to MIRO like oscillations of the absorbed microwave power, they are related to consecutive
collisions between an electron and impurity (memory effects) \cite{IvanPow,Zhirov,Dyakonov}. 
As expected at small amplitudes of microwave field the absorbed power ${\cal P}(J)$ is proportional to ${E_{ac}}^2$ (see Fig.S4 in SI Sec.6 and 
discussion of parameter scaling dependence given there; details of numerical computations of absorption power are given in SI Sec.5.). 
For the repulsive impurity the results of the quantum master equation and the classical kinetic theory are in quantitative agreement, the agreement is less accurate for attractive impurities but the correct peak positions are reproduced by the kinetic equation even in this case. It is likely that the quantum dynamics inside an attractive impurity has stronger deviations from the classical behavior when its depth is comparable with the Fermi energy $E_F$.
The good overall agreement between quantum and classical simulations holds even if the impurity radius $r_d$ is only two times larger than the magnetic length, in supplementary information we shows that this trends continues for even smaller $r_d$. 
An interesting property that appears in Fig.~\ref{fig2} is that the oscillations in microwave absorption are phase shifted between the repulsive and attractive potentials, this shift is related to the additional $\pi$ contribution that appears in Eq.~(\ref{eq:map}) for the attractive potential. According to this equation the change of orbital momentum due to the action of the microwave field over a cyclotron period vanishes for integer values of $J$. This leads to an absorption dip 
at integer $J$ values with different lineshapes depending on the sign of the potential of the impurity. For a repulsive potential (at $\omega_c \tau = 100$) integer $J$ correspond to an absorption minimum with maxima occurring close to half integer $J$ values. For the attractive potential, the phase $\pi$ shift in the map due to the interaction sign tends to move the position of the maxima near integer values which leads to a characteristic lineshape with a double peak structure centered around integer values. We note that earlier calculations of the absorbed power in the MIRO regime for the case of a repulsive potential gave similar results \cite{IvanPow} for the position of absorption peaks/dips, Eq.~(\ref{eq:map}) allows to generalize these results to the case of an arbitrary electron/impurity interaction.  
We also note that the results for the absorption dependence on $J$ obtained from  the kinetic equation
are also well reproduced by the symplectic map description (\ref{eq:map}) (see Fig.S2 in SI Sec.5). 
The dependence ${\cal P}(J)$ is not sensitive to variation of $r_d/\ell_B$ as long as $r_d$ is larger than $\ell_B$, in the limit $r_d \le \ell_B$ additional quantum oscillations appear around the average semi-classical absorption curve but those will probably ensemble average to zero in a macroscopic sample (see Fig.S3 in SI Sec.5). 

\subsection{V Charge density distribution induced by microwave field}

The absorbed power ${\cal P}$ from the external microwave field creates a charge redistribution in a around
of impurity. To see this it is convenient to present this power ${\cal P}({E}_{ac})$ in the following form
(see also SI Sec.5):
\begin{align}
{\cal P} = {\rm Tr} ({\hat \rho} \mathbf{v}\cdot q\mathbf{E}_{ac} ) 
\approx \omega {\rm Tr} ({\hat \rho} \mathbf{r})  \cdot ( \mathbf{e}_z \times q \mathbf{E}_{ac} ) \; .
\label{eq:power}
\end{align}
This formula clearly shows that in the rotating frame 
there is an appearance of a stationary dipole moment of charge induced by $\mathbf{E}_{ac}$.
Thus in the laboratory frame we have a rotating dipole which creates a correction
 to the field acting on electrons.

\begin{figure}[h]
\centering
\input{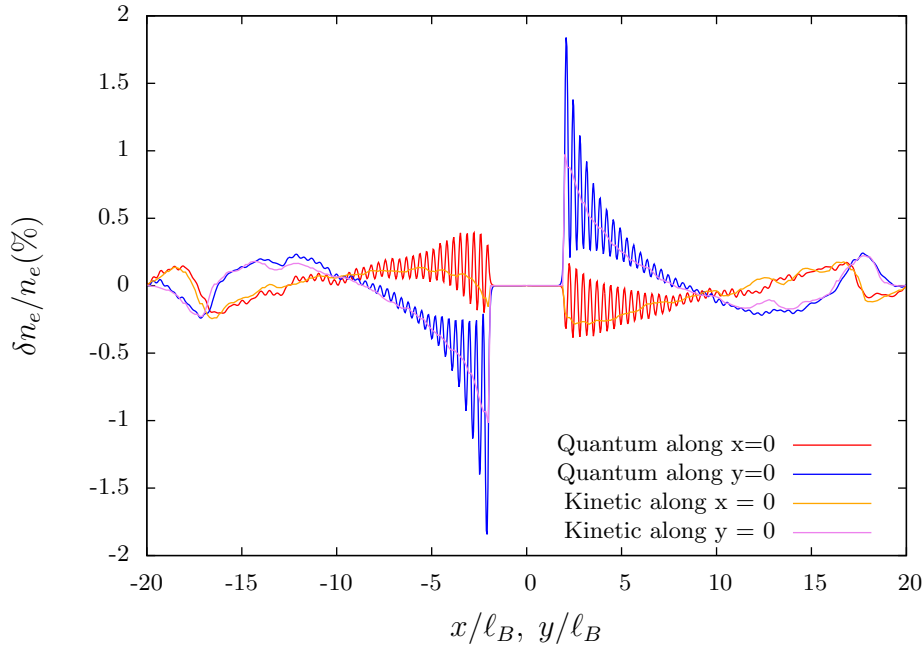}
\caption{\label{fig3}
Charge density $\rho(x,y)$ dependence on $x$ (at $y=0$)
and on $y$ (at $x=0$) obtained from the quantum master equation and
the classical kinetic equation presented in the rotation frame.
Here the system parameters are the same as
for the repulsive case in Fig.~\ref{fig2} with $J = 2.7$ for $r_d/\ell_B=2$ at
$qE_{ac}\ell_B/\hbar \omega_c=0.1$ and $\epsilon_F/\hbar \omega_c=40$ with a relaxation time $\omega_c \tau=10$.
}
\end{figure}
\begin{figure}[h]
\includegraphics[width=0.450\columnwidth]{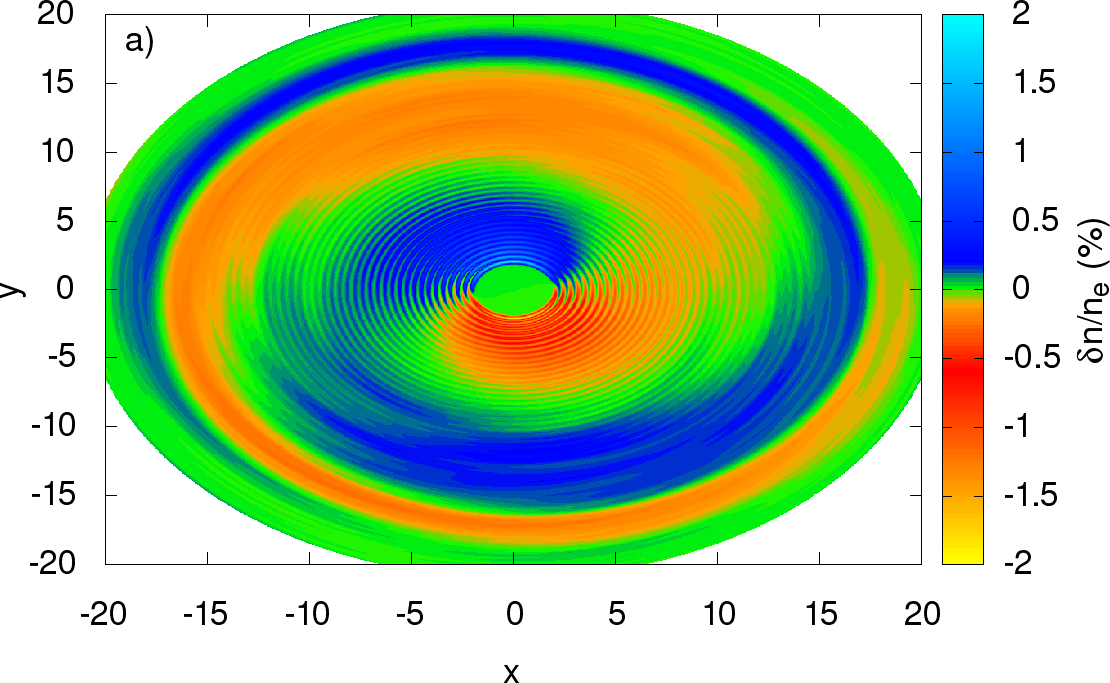}
\includegraphics[width=0.450\columnwidth]{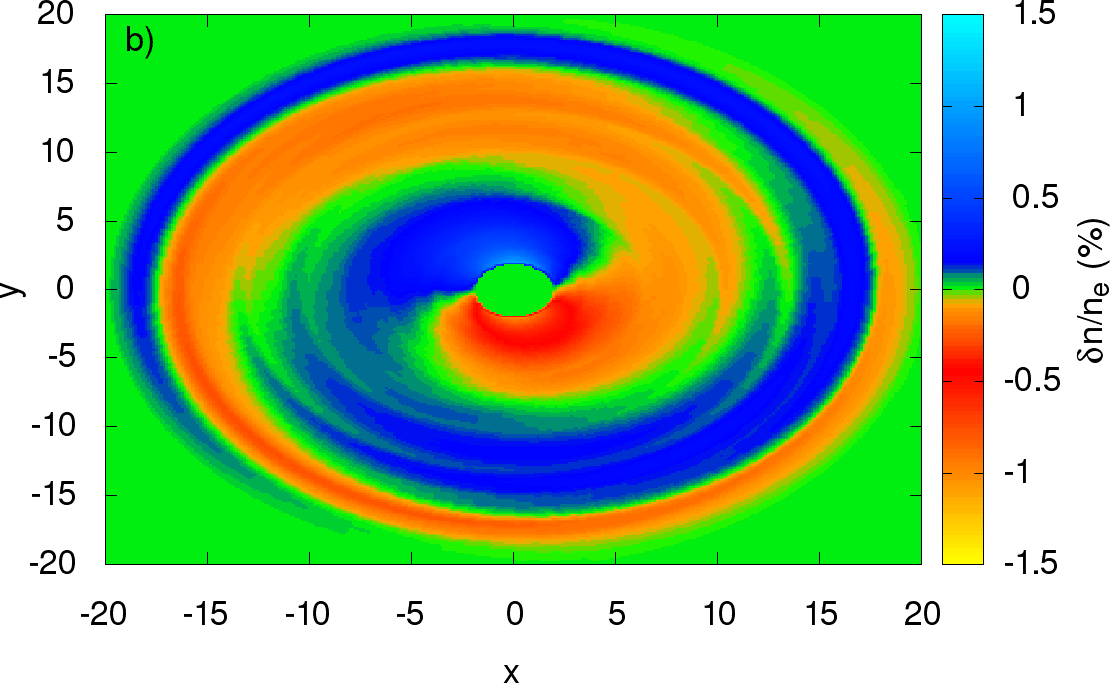}
\caption{\label{fig4}
Relative charge density variation $\rho(x,y) = \delta n /n_0$ (expressed in percent)
shown in $(x,y)$ plane in the rotation frame where a microwave field is directed alone
$x-$axis. Left panel shows the results of quantum simulations with the master equation,
left panel show the results of the classical kinetic equation.
The parameters are the same as in Fig.~\ref{fig3}.
}
\end{figure}

The stationary solutions of the master equation  (\ref{eq:masterstat})
and the kinetic equation (\ref{eq:kinetic})
allow to compute the charge density variation induced by  $\mathbf{E}_{ac}$.
The charge density distribution in presence of microwave field is shown as a function of  $x$ and $y$ coordinates in the rotating frame 
in Fig.~\ref{fig3}. In the vicinity of the impurity
we have strong quantum Friedel like oscillations of density \cite{Friedel}. 
The average density lineshape from the master equation (\ref{eq:master1}) 
calculation is well approximated by results 
from the classical kinetic equation (\ref{eq:kinetic}). We note that the amplitude of  
density variation increases near the impurity in both cases reaching rather high values 
of 1-2 percent of the total electron density. The steady-state relative variation 
of charge density $\delta n /n_0$ in the whole $(x,y)$ plane is shown in the rotating frame in Fig.~\ref{fig4} 
revealing the formation of a charge density vortex. 
Since this structure is accurately reproduced from the classical dynamics, we attribute it to 
a nonlinear resonance between the microwave frequency and 
collision frequency with the impurity
which is well described by the Poincar\'e section of Fig.~\ref{fig1}. 

The results of Fig.~\ref{fig3} show that in the laboratory
frame microwave driving creates a rotating charge density vortex
that forms a rotating dipole moment.
This dipole moment will create a rotating electric field that 
leads to a renormalization of the external field in the vicinity of the impurity.
We now address the issue of the calculation of this renormalized field from the rotating density profiles. 

\subsection{VI MIRO of short range field near impurity}

In the absence of screening, the rotating charge density vortex $\delta n(\mathbf{r})$ 
around the impurity creates an electric field that is simply given by the Coulomb integral 
\begin{align}
\mathbf{E}_{ee}(\mathbf{r}) = \frac{q}{4 \pi \epsilon_0 \epsilon_r} \int \delta n(\mathbf{r}') 
\frac{ \mathbf{r} - \mathbf{r}' }{| \mathbf{r} - \mathbf{r}' |^3 } d^2 r' 
\label{eq:fieldcc}
\end{align}

Within a weakly-interacting electron approximation, this corresponds to 
a Hartree type contribution to the rescaled external field. We developed our Floquet theory for 
non interacting electrons it is thus difficult to estimate the exact form of the interaction related corrections. 
Qualitatively the strength of this Coulomb field may be reduced by screening from other electrons, 
however the characteristic length-scale of the rotating charge density vortex is of the order of 
the cyclotron radius. A full-screening scenario seems thus unlikely as it is difficult to justify 
how the electron gas could provide a complete screening on this length-scale at a frequency $\omega$ 
which is several times larger than $\omega_c$, we thus neglect screening effects here and 
we present numerical results for the unscreened Coulomb integral given by Eq.~(\ref{eq:fieldcc}). 

\begin{figure}[h]
\includegraphics[width=0.6\columnwidth]{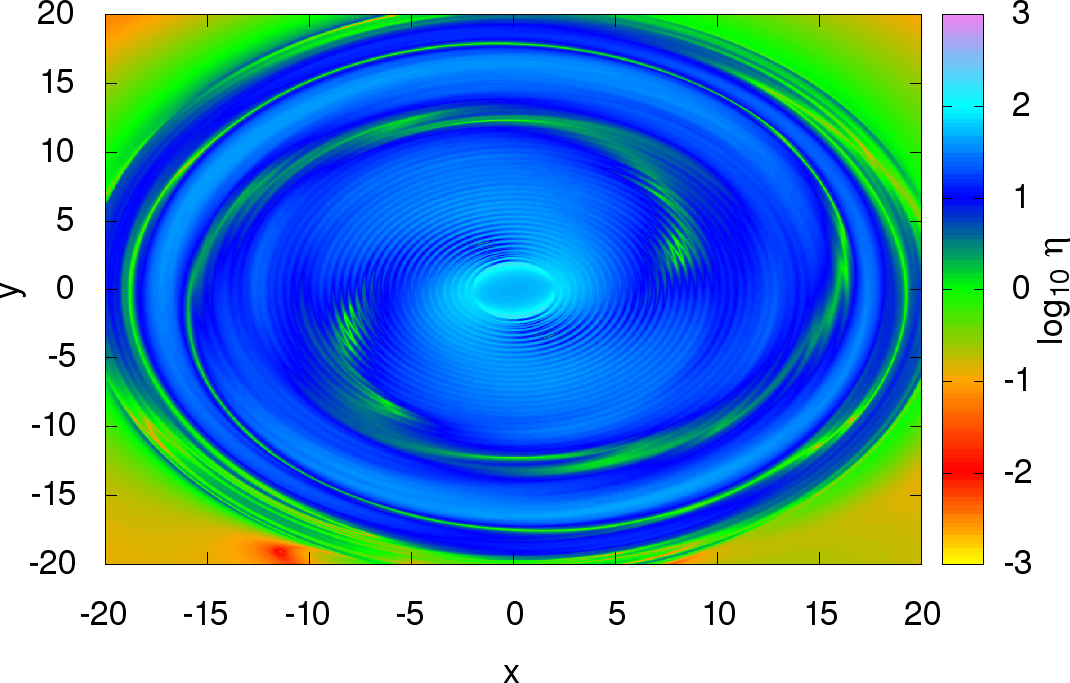}
\caption{\label{fig5}
Variation of the enhancement factor
$E_{ee}/E_{ac}$ of the amplitude of effective field acting on 
an electron due to the density variation 
in $(x,y)$ plane (see Fig.~\ref{fig4} left panel)
induced by the external microwave field.
The results, shown in the rotating frame, 
are obtained from the quantum master equation
and the relations given by Eq.(\ref{eq:fieldcc}).
}
\end{figure}

To characterize the strength of the field renormalization we introduce the enhancement factor $\eta = E_{ee}/E_{ac}$, which is the ratio between the amplitude of the correction field from Eq.~(\ref{eq:fieldcc}) to the amplitude of the external driving $E_{ac}$. The logarithm of this enhancement factor is shown on Fig.~\ref{fig5} on $(x,y)$ plane using the data for the master equation density distribution from Fig.~\ref{fig4}.
These results clearly show that the field can be enhanced by an order of magnitude in the vicinity of impurity, while far from the impurity the enhancement factor goes to zero as expected. The strongest enhancement is obtained at the boundary of the impurity and reaches two orders of magnitude.

Checking the parametric dependence of the rotating density vortex in the semi-classical model, we find that $\delta n(r)$ depends only weakly on $r_d/R_c$ and $\omega \tau$ in a broad parameter range. This allows us to estimate the typical value of the enhancement as $\eta = E_{ee}/E_{ac} \sim q^2 n_e / (m \omega \epsilon v_F) \simeq 6$ for typical values $\omega = 2\pi \times 100{\rm GHz}$, $n_e = 3.5 \times 10^{11}\;{\rm cm^{-2}}$, and dielectric constant $\epsilon_d = 10 \epsilon_0$. The obtained order of magnitude is consistent with the values of $\eta$ in Fig.~\ref{fig4} at a distance of around a cyclotron radius away from the impurity, an even larger enhancement enhancement occurs at smaller distances. The estimate for $\eta$ can be cast in a form that highlights its dependence on $r_s$ the usual interaction strength parameter in 2DEG  $\eta \sim r_s E_F / \hbar \omega$ (we remind that $r_s = (\pi n_e a_0^2)^{-1/2}$ where $a_0 = \frac{4 \pi \epsilon_d \hbar^2}{m q^2}$ is the Bohr radius, and $q$ is the electron charge). A detailed discussion of scaling dependence on system parameters is given in SI Sec.6 with Fig.S4, Fig.S5, Fig.S6, Fig.S7.

The large values of the field enhancement parameter $\eta$ show that a full interacting electron calculation is required to obtain consistently the field acting on the electrons. Our results suggest however the following qualitative picture: the external microwave field $E_{ac}$ generates a strong effective circular polarized microwave
field in the vicinity of the impurity which is significantly larger than the external field $E_{eff} \gg E_{ac}$. This field $E_{eff}$  is maximal at $r = r_d$ 
and decays to the external field on a typical scale $r_{eff} \lesssim R_c$.
It has a the same circular polarization as the driving microwave field since it is created by a 
charge density vortex that rotates at the microwave frequency $\omega$.

\begin{figure}[h]
\includegraphics[width=0.450\columnwidth]{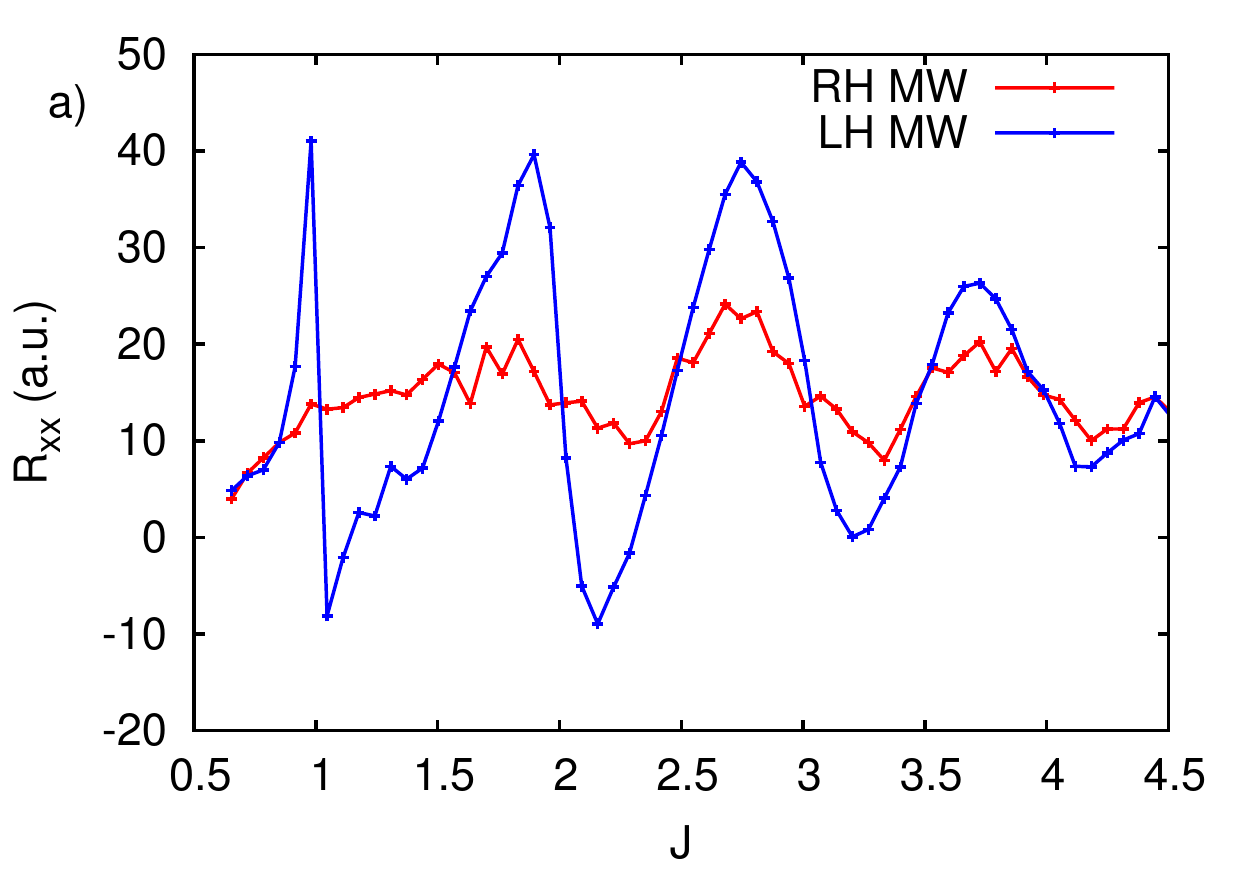}
\includegraphics[width=0.450\columnwidth]{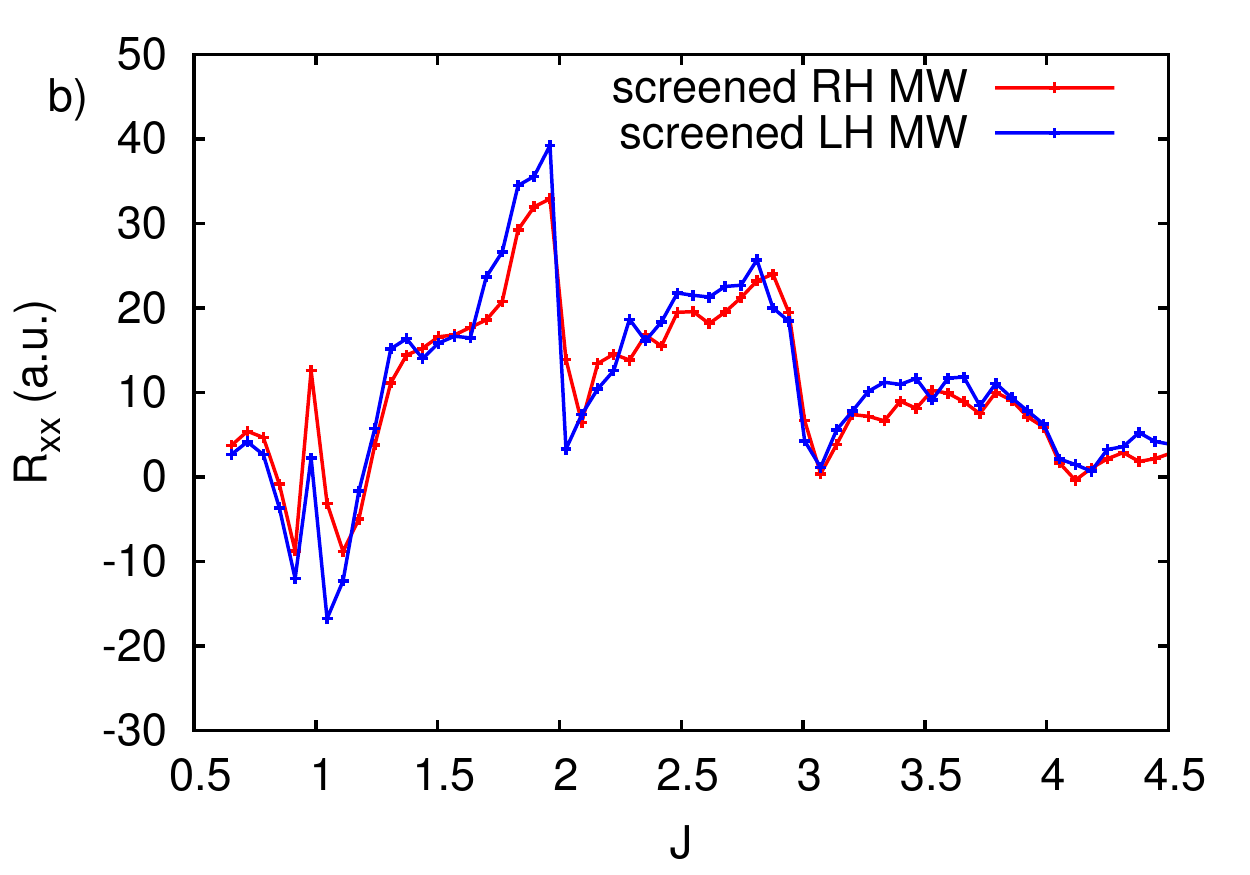}
\caption{\label{fig6}
Dependence of resistivity $R_{xx}$ (in arbitrary units a.u.)
on $J=\omega/\omega_c$. Left panel: the case of initial external microwave field; right panel: the case of screened renormalized field 
acting on a length-scale $r_{eff}/R_c =0.3$.
The results are obtained from the numerical simulation of classical
Hamiltonian dynamics of electrons in presence of impurities with Gaussian potential $U_{amp} \exp(-r^2 r_d^{-2})$ with a density $n_i r_d^2 = 10^{-3}$ an amplitude $U_{amp}/\epsilon_F = 1.5$ and a characteristic radius $r_d \omega/v_F = 0.1$.
}
\end{figure}

We are thus justified to consider a model where the driving microwave field 
is localized around the impurity instead of being homogeneous in space as considered so far in most models 
(with the exception of work \cite{Mikhailov} which also considered a strongly inhomogeneous field 
but only close to metallic contacts). We show through numerical simulations, that taking 
into account the non-homogeneity of the microwave field can indeed solve the puzzle of 
the polarization independence \cite{Smet,Kvon}. For this purpose we use the approach 
of \cite{Dyakonov} computing the resistivity $R_{xx}$
for a sample of finite size with a fixed density of impurities
using the Hamiltonian dynamics for electrons.
The details of these simulations are given in SI Sec.6.
In Fig.~\ref{fig6} we compare 
the dependence of resistivity $R_{xx}$,
on $J$ for the case of microwave field homogeneous
in the whole space $\mathbf{E}_{ac}$ (left panel) and for the case of a model screened microwave
field $\mathbf{E}_{ac} \exp(-r^2 r_{eff}^{-2})$ which decays with a Gaussian profile as function of the distance from the impurity 
with a characteristic range $r_{eff} \approx R_c/5$ .
The main result of these simulations is a qualitatively
different dependence of $R_{xx}$
on positive and negative $J$ values.
For a microwave field homogeneous in space there is a
strong asymmetry between positive and negative
circular polarization. In contrast,
to a screened microwave field
localized in a vicinity of impurity
in a range $r_{eff} \sim r_d \ll R_c$
there is no dependence of the sign of polarization
(sign of $J$) in agreement with the experimental 
results reported in \cite{Smet,Kvon}.


The origin of the absence of sign dependence on $J$
for a localized field is rather simple:
the range of the field is much smaller then the cyclotron radius
($r_{eff} \sim r_d \ll R_c$)
so that the energy change takes place only near impurity
and the cyclotron resonance appearing in the kick amplitude
$F \propto 1/(\omega-\omega_c)$ 
in (\ref{eq:map}) is absent.
In fact the kick amplitude is determined
only by a time interval $\Delta t_{eff} = r_{eff}/v_F$
of interaction  of change with the field $E_{eff}$  
near impurity. It can be estimated as in (\ref{eq:map}) with
$F \approx q E_{eff} r_{eff}/v_F$ being independent
of the sign of polarization. In a certain sense
we obtain a screened field near impurity.
This situation is similar to the Azbel'-Kaner effect for cyclotron resonance absorption in metals. Due to the skin effect, the microwave field is screened in the bulk of the metal on a length scale much smaller than the cyclotron radius. The electron energy change induced by
the interaction with the microwave field then takes place only in a vicinity of the metal surface \cite{Kaner,Azbel}. In this case cyclotron resonance occurs not only at the cyclotron frequency but at integer harmonics (similarly to MIRO) and there is no polarization dependence since the interaction range is much smaller
than the cyclotron radius.

\begin{figure}[h]
\includegraphics[width=0.450\columnwidth]{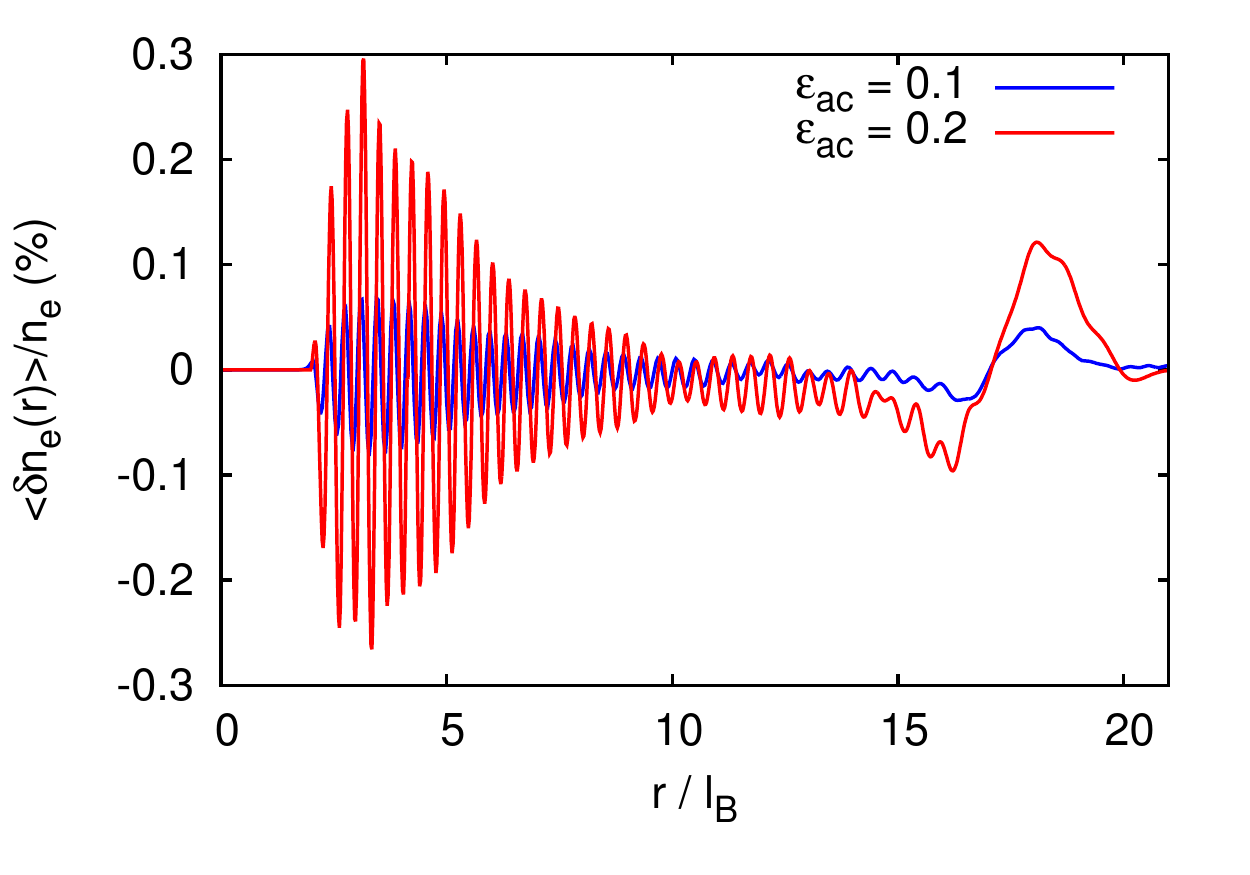}
\caption{\label{fig7}
Relative variation of the time averaged electron charge 
density $<\delta n_e(r)>/n_e$ induced by a microwave irradiation, 
as a function of radial distance $r$ from the impurity.
The results are shown two microwave fields $\epsilon_{ac} = 0.1$ and  $0.2$ ($\epsilon_{ac} = q E_{ac} \ell_B/ \hbar \omega_c$).
Other parameters are as in Fig.~\ref{fig3} ($r_d/\ell_B=2$, $E_F/\hbar \omega_c=40$).
$\omega_c \tau=100$ 
}
\end{figure}

We have shown that the rotating charge density vortex around the impurity strongly renormalizes the external microwave field since it rotates at the frequency of the external excitation. It is interesting to know if a renormalization of the static potential created by the impurity is present as well. In a mean field Hartee type approximation, such a renormalization could be created by the time averaged density profile of the rotating vortex. The time average in the laboratory frame corresponds to an average over the polar angle in the rotating frame. For the semi-classical calculation in Fig.~\ref{fig4}.b) the charge redistribution $\delta n_e$ seems antisymmetric with respect to the origin and numerical integration over angles gives a vanishing result within numerical accuracy. The situation is different in the quantum case where a closed form analytic expression for the angle averaged density can be derived (see supplementary materials). The time (angle) averaged density distribution  $<\delta n_e(r)>$ obtained from the quantum calculation for the parameters of Fig.~\ref{fig4}.a) is shown on Fig.~\ref{fig7}, while the time average value of $<\delta n_e(r)>/n_e$ are smaller compared to the typical values for the charge density vortex they grow with microwave power. Compared to regular equilibrium Friedel oscillations that decay on the scale of the Fermi-wavelength, the time averaged density profile $<\delta n_e(r)>/n_e$ induced by microwave irradiation spreads over a much larger scale given by the cyclotron radius. In principle it is thus possible to enter a regime where the collision cross section of the impurity is mainly determined by the long range electrostatic potential created by $<\delta n_e(r)>$ rather that by the bare short range potential of the impurity. A quantitative treatment of this scenario is beyond the scope of this work where electron-electron interactions are not taken into account at the quantum level. In \cite{Zhirov} we suggested an alternative scenario to the macroscopic domain formation for ZRS in semiconducting hetero-structures, where the charge redistribution around an isolated sharp impurity would create a smooth cloaking potential around the impurity allowing an adiabatic passage around the impurity in which the momentum scattering would be strongly suppressed. An advantage of such a scenario compared to the more conventional macroscopic domain formation is that charges do not have to overcome macroscopic distances to create the electric field domains since the charge redistribution takes place on the scale of $R_c$ around short range impurities. These calculations thus show that this scenario is plausible within the non-interacting electrons quantum model. 

\subsection{VII Discussion}

We preformed extensive numerical simulations
of electron dynamics in a vicinity of
impurity in presence of a 
circular polarized microwave irradiation.
Our studies demonstrate that a description 
based on the classical Newton dynamics 
provides a good approximate description
being close to the exact solution 
of the Schr\"odinger equation
in the regime when the size of impurity 
$r_d$ is larger than
the magnetic length $\ell_B$
and the Fermi energy contains
many quanta of cyclotron frequency
($E_F \gg \hbar \omega_c$).
We also show that the classical dynamics is
well described by the collision symplectic map
which allows to understand the origin
of periodic dependence of MIRO on 
the frequency ratio $J=\omega/\omega_c$
and disappearance of microwave effect
at integer $J$ values.

The absorption power of charges 
in an impurity vicinity 
is obtained from the numerical
solution of quantum master equation
and the classical kinetic, or Vlasov, equation
which have  approximately
similar average dependence on
system parameters, a part of Friedel like
oscillations present in the quantum case.
The results obtained from the master and kinetic equations
show the emergence of a strong charge redistribution
created by a microwave irradiation taking the form of a rotating charge density vortex. Thus the irradiation
generates a rotating dipole field localized 
near impurity. The strength of this induced localized field
is significantly larger than the external microwave field,
enhancing the external field near the impurity.
This strong inhomogeneity of the microwave field results in absence of polarization dependence
and absence of the cyclotron resonance 
in agreement with the experimental results reported in \cite{Smet,Kvon}.
This effect is similar to the Azbel'-Kaner
effect \cite{Kaner,Azbel} for resistance magneto-oscillation
induced microwave screened in a vicinity
of metal surface.

At the final stage of this work there appeared the MIRO results 
for electrons on a liquid helium under circularly polarized microwave excitation \cite{heliumcr2017}.
These  experimental results show that the resistivity changes significantly
with the sign of polarization in contrast to the results for 2DEG in GaAs \cite{Smet,Kvon,Ganichev}
where MIRO are almost independent on the sign of the circular polarization. 
Since the electrons densities for electrons on helium 
are several orders of magnitude smaller than in GaAs 2DEG, 
the field amplification effects discussed here are expected to be negligible. 
The difference in circular polarization dependence between the two systems could 
thus be a hint of the importance of external field renormalisation in high density electron systems. 

\section{References}


\begin{thebibliography}{10}

\bibitem{Hall1} von Klitzing, K., Dorda, G. \& Pepper, M.
\newblock {\it New method for high-accuracy determination of 
the fine-structure constant based on quantized hall resistance.}
\newblock {\em Phys. Rev. Lett.} {\bf 45}, 494--497 (1980).

\bibitem{Hall2} Tsui, D.C., Stormer, H.L. \& Gossard, A.C.
\newblock {\it Two-dimensional magnetotransport in the extreme quantum limit.}
\newblock {\em Phys. Rev. Lett.} {\bf 48}, 1559--1562 (1982).

\bibitem{Demler} Kitagawa, T., Berg, E., Rudner, M. \& Demler, E.
\newblock {\it Topological characterization of periodically driven quantum systems.}
\newblock {\em Phys. Rev. B} {\bf 82}, 235114 (2010).

\bibitem{Galitski} Lindner, N.H., Refael, G. \& Galitski, V.
\newblock {\it Floquet topological insulator in semiconductor quantum wells.}
\newblock {\em Nature Phys.} {\bf 7}, 490--495 (2011).

\bibitem{Gawędzki} D. Carpentier, P. Delplace, M. Fruchart, and K. Gawędzki
\newblock {\it Topological Index for Periodically Driven Time-Reversal Invariant 2D Systems}
\newblock {\em Phys. Rev. Lett.} {\bf 114}, 106806 (2015)

\bibitem{Segev1} Rechtsman, M.C. \& {\it et al.}.
\newblock {\it Photonic Floquet topological insulators.}
\newblock {\em Nature} {\bf 496}, 196--200 (2013).

\bibitem{Wang} Wang, Y.H., Steinberg, H., Jarillo-Herrero, P. \& Gedik, N.
\newblock {\it Observation of Floquet-Bloch states on the surface of a topological insulator.}
\newblock {\em Science} {\bf 342}, 453--457 (2013).

\bibitem{Segev2} Weimann, S. \& {\it et al.}.
\newblock {\it Topologically protected bound states in photonic parity–time-symmetric crystals.}
\newblock {\em Nature Mat.} doi:10.1038/nmat4811 (2016).

\bibitem{Zudov1} Zudov, M.A., Du, R.R., Simmons, J.A. \& Reno, J.L.
\newblock {\it Shubnikov–de Haas-like oscillations in millimeterwave photoconductivity 
in a high-mobility two-dimensional electron gas.}
\newblock {\em Phys. Rev. B} {\bf 64}, 201311(R) (2001).

\bibitem{Mani} Mani, R.G. \& {\it et al.}.
\newblock {\it Zero-resistance states induced by
electromagnetic-wave excitation in GaAs/AlGaAs heterostructures.}
\newblock {\em Nature} {\bf 420}, 646--650 (2002).

\bibitem{Zudov2} Zudov, M.A., Du, R.R., Pfeiffer, L.N. \& West, K.W.
\newblock {\it Evidence for a new dissipationless effect in 2D electronic transport.}
\newblock {\em Phys. Rev. Lett.} {\bf 90}, 045807 (2003).

\bibitem{Denis1} D. Konstantinov and K. Kono
\newblock {\it Novel Radiation-Induced Magnetoresistance Oscillations in a Nondegenerate Two-Dimensional Electron System on Liquid Helium}
\newblock {\em Phys. Rev. Lett.} {\bf 103}, 266808 (2009)

\bibitem{Bykov} A. A. Bykov, 
\newblock {\it Microwave-induced magnetic field state with zero conductivity in GaAs/AlAs Corbino disks and hall bars}
\newblock JETP Lett. 87, 551 (2008).

\bibitem{ZudovGe} M. A. Zudov, O. A. Mironov, Q. A. Ebner, P. D. Martin, Q. Shi, \& D. R. Leadley
\newblock {\it Observation of microwave-induced resistance oscillations in a high-mobility two-dimensional hole gas in a strained Ge/SiGe quantum well}
\newblock {\em Phys. Rev. B} {\bf 89}, 125401 (2014).

\bibitem{ZnO} D. F. Kärcher, A. V. Shchepetilnikov, Yu. A. Nefyodov, J. Falson, I. A. Dmitriev, Y. Kozuka, D. Maryenko, A. Tsukazaki, S. I. Dorozhkin, I. V. Kukushkin, M. Kawasaki, and J. H. Smet
\newblock {\it Observation of microwave induced resistance and photovoltage oscillations in MgZnO/ZnO heterostructures}
\newblock {\em Phys. Rev. B} {\bf 93}, 041410(R) (2016).

\bibitem{Denis2} R. Yamashiro, L. V. Abdurakhimov, A. O. Badrutdinov, Yu. P. Monarkha, and D. Konstantinov
\newblock {\it Photoconductivity Response at Cyclotron-Resonance Harmonics in a Nondegenerate Two-Dimensional Electron Gas on Liquid Helium}
\newblock {\em Phys. Rev. Lett.} {\bf 115}, 256802 (2015).

\bibitem{Natcom} Chepelianskii, A.D., Watanabe, M., Nasyedkin, K., Kono, K. \& 
Konstantinov, D.
\newblock {\it An incompressible state of a photo-excited electron gas.}
\newblock {\em Nature Comm.} {\bf 6}, 7210 (2015).

\bibitem{Theory1} V.I. Ryzhii, Sov. Phys. Solid State 11, 2078 (1970).

\bibitem{Girvin} A. C. Durst, S. Sachdev, N. Read, and S. M. Girvin, 
\newblock {\it Radiation-induced magnetoresistance oscillations in a 2D electron gas}
\newblock Phys. Rev. Lett. {\bf 91}, 086803 (2003).

\bibitem{TheoryIvan1} I. A. Dmitriev, M. G. Vavilov, I. L. Aleiner, A. D. Mirlin, and D. G. Polyakov
\newblock {\it Theory of microwave-induced oscillations in the magnetoconductivity of a two-dimensional electron gas}
\newblock {\em Phys. Rev. B} {\bf 71}, 115316 (2005).

\bibitem{TheoryIvan2} I. A. Dmitriev, A. D. Mirlin, and D. G. Polyakov
\newblock {\it Theory of Fractional Microwave-Induced Resistance Oscillations}
\newblock {\em Phys. Rev. Lett.}  {\bf 99}, 206805 (2007)

\bibitem{Zudovrmp} Dmitriev, I.A., Mirlin, A.D., Polyakov, D.G.  \& Zudov, M.A.
\newblock {\it Nonequilibrium phenomena in high Landau levels.}
\newblock {\em Rev. Mod. Phys.} {\bf 84}, 1709-1763 (2012).

\bibitem{Zhirov} Zhirov, O.V., Chepelianksii, A.D.  \& Shepelyansky, D.L.
\newblock {\it Towards a synchronization theory of microwave-induced zero-resistance states.}
\newblock {\em Phys. Rev. B} {\bf 88}, 035410 (2013).

\bibitem{Dyakonov} Beltukov, Y.M.  \& Dyakonov, M.I.
\newblock {\it Microwave-induced resistance oscillations as a classical memory effect.}
\newblock {\em Phys. Rev. Lett.} {\bf 116}, 176801 (2016).

\bibitem{Smet} Smet, J.H.  \& {\it et al.}.
\newblock {\it Circular-polarization-dependent study of the microwave photoconductivity
in a two-dimensional rlectron system.}
\newblock {\em Phys. Rev. Lett.} {\bf 95}, 116804 (2005).

\bibitem{Mani1} R. G. Mani, A. N. Ramanayak, and W. Wegscheider 
\newblock {\it Observation of linear-polarization-sensitivity in the microwave-radiation-induced 
magnetoresistance oscillations}, 
\newblock {\em Phys. Rev. B} {\bf 84}, 085308 (2011)

\bibitem{Mani2} A. N. Ramanayaka, R. G. Mani, J. Inarrea, and W. Wegscheider
\newblock {\it Effect of rotation of the polarization of linearly polarized microwaves on the radiation-induced magnetoresistance oscillations}
\newblock {\em Phys. Rev. B} {\bf 85}, 205315 (2012)

\bibitem{Kvon} Herrmann, T.  \& {\it et al.}
\newblock {\it Analog of microwave-induced resistance oscillations induced 
in GaAs heterostructures by terahertz radiation.}
\newblock {\em Phys. Rev. B} {\bf 94}, 081301(R) (2016).

\bibitem{Ganichev} Herrmann, T.  \& {\it et al.}
\newblock {\it Magnetoresistance oscillations induced by high-intensity terahertz radiation.}
\newblock {\em Phys. Rev. B} {\bf 96}, 115449 (2017).

\bibitem{Chepelianskii} Chepelianskii, A.D.  \& Shepelyansky, D.L.
\newblock {\it Microwave stabilization of edge transport and zero-resistance states.}
\newblock {\em Phys. Rev. B}, {\bf 80}, 241308(R) (2009).

\bibitem{Mikhailov} S. A. Mikhailov
\newblock {\it Theory of microwave-induced zero-resistance states in two-dimensional electron systems}
\newblock {\em Phys. Rev. B}, {\bf 83}, 155303 (2011)

\bibitem{Kaner} Kaner, E.A.  \& Azbel', M.I.
\newblock {\it Theory of cyclotron resonance in metals.}
\newblock {\em Sov. Phys. JETP} {\bf 6(33)}, 1126--1134 (1958) 
[J. Exptl. Theoret. Phys. U.S.S.R {\bf 33}, 1461--1471 (1957)].

\bibitem{Azbel} Azbel', M.Y.  \& Kaner, E.A.
\newblock {\it Cyclotron resonance in metals.}
\newblock {\em J. Phys. Chem. Solids (Pergamon Press)} {\bf 6}, 113--135 (1958).

\bibitem{Kittel} Kittel, C. 
\newblock {\it Quantum theory of solids.}
\newblock {\em Willey, New York} (1963).

\bibitem{Haake} Haake, F. 
\newblock {\it Quantum sibnatures of chaos.}
\newblock {\em Springer-Verlag, Berlin} (2001).

\bibitem{Husimi1} Crespi, B., Perez, G.  \& Chang, S-J.
\newblock {\it Quantum Poincar\'e sections for two-dimensional billiards.}
\newblock {\em Phys. Rev. E}, {\bf 47}, 986--991 (1993).

\bibitem{Husimi2} Hentschel, M., Schomerus, H.  \& Schubert, R.
\newblock {\it Husimi functions at dielectric interfaces: Inside-outside
duality for optical systems and beyond.}
\newblock {\em Europhys. Lett.}, {\bf 62(5)}, 636--642 (2003).

\bibitem{Lichtenberg}  Lichtenberg, A. \&  Lieberman M.
\newblock {\it Regular and chaotic dynamics.}
\newblock {\em Springer-Verlag, New York} (1992).

\bibitem{Chirikov} Chirikov, B.V.
\newblock {\it A universal instability of many-dimensional oscillator systems.}
\newblock {\em Phys. Rep.}, {\bf }, 263--379 (1979).

\bibitem{Kohler} Kohler. S., J\"org L., P. H\"anggi 
\newblock {\it Driven quantum transport on the nanoscale.}
\newblock Physics Reports {\bf 406} (2005), 379

\bibitem{Hanggi} Grifoni M., H\"anggi P.
\newblock {\it Driven quantum tunneling.}
\newblock Physics Reports {\bf 304} (1998), 229

\bibitem{IvanPow} I.A. Dmitriev, A.D. Mirlin and D.G. Polyakov
\newblock {\it Oscillatory ac conductivity and photoconductivity of a two-dimensional electron gas:
Quasiclassical transport beyond the Boltzmann equation}
\newblock Phys. Rev. B {\bf 70}, 165305 (2004)

\bibitem{Friedel} Friedel, J.
\newblock {\it Metallic alloys.}
\newblock {\em J. Nuovo Cimento}, {\bf 7 (Sup. 2)}, 287--311 (1958).

\bibitem{heliumcr2017} Zadorozhko, A.A., Monarkha Yu.A., Konstantinov D.
\newblock {\it Circular-polarizetion-dependent study of microwave-induced conductivity oscillations 
in a two-dimensional electron gas on liquid helium.}
\newblock {\em arXiv:1710.00531[cond-mat.mes-hall]} (2017).

\end{thebibliography}

\section*{Acknowledgments}

We thank O.V.Zhirov for discussions and help
in construction of a quantum Husimi function.
ADC was supported by the France ANR project SPINEX,
DLS was supported in part by the Pogramme Investissements
d'Avenir ANR-11-IDEX-0002-02, reference ANR-10-LABX-0037-NEXT (project THETRACOM).




%
%
%
%


%
%
%



%

\begin{widetext}

\section*{Supplementary Information}
\setcounter{equation}{0}
\renewcommand{\theequation}{S\arabic{equation}}
\setcounter{figure}{0}
\renewcommand\thefigure{S\arabic{figure}}
\renewcommand{\figurename}{Supplementary Figure}

\section*{1. Numerical solution of quantum master equation}
\label{sisec1}

\subsection{Stationary Schr\"odinger equation without microwaves}

The classical Lagrangian/Hamiltonian read : 
\begin{align}
L  &= \frac{m}{2} \left( \dot{r}^2 + r^2 \dot{\theta}^2 \right) - \frac{m \omega_c}{2} r^2 \dot{\theta} - U_w(r)\\
H_0 &= \frac{p_r^2}{2 m} + \frac{1}{2 m r^2} \left(p_\theta + \frac{m \omega_c r^2}{2}  \right)^2 + U_w(r) \\ 
&= \frac{p_r^2}{2 m} + \frac{p_\theta^2}{2 m r^2} + \frac{\omega_c}{2} p_\theta + \frac{m \omega_c^2 r^2}{8}  + U_w(r) 
\end{align}

where $r, \theta$ are polar coordinates, $U_w(r)$ is the impurity potential and:
\begin{align}
p_\theta = m r^2 \dot{\theta} - \frac{m \omega_c r^2}{2} 
\end{align}

In absence of disc potential, the angular momentum can be related with the geometrical parameters 
of the trajectory: 
\begin{align}
p_\theta = m \omega_c \frac{R_L^2 - L_c^2}{2}
\end{align}
where $R_L$ is the Larmor radius and $L_c$ is the distance from the guiding center to the coordinates origin.

We note that we can not obtain the Schr\"odinger equation directly from quantization of this Hamiltonian, 
this seems due to the ill-defined nature of the operator ${\hat p}_r$ in two dimensions \cite{sh2d}. Instead 
the eigenvalue equation takes the following form which comes from rewriting the Schroedinger equation in cylindrical coordinates \cite{LandauT3}: 
\begin{align}
{\hat H}_0 \psi &= -\frac{\hbar^2}{2 m} \left[ \frac{1}{r} \frac{\partial}{\partial r} \left( r \frac{\partial \psi}{\partial r} \right) + \frac{1}{r^2} \frac{\partial^2 \psi}{\partial \theta^2} \right] - \frac{i \hbar \omega_c}{2} \frac{\partial \psi}{\partial \theta} + \frac{m \omega_c^2 r^2}{8} \psi + U_w(r) \psi\\ 
&= E \psi 
\end{align}

We introduce 
\begin{align}
\psi(r,\theta) = \frac{\chi(r)}{\sqrt{r}} \frac{e^{i L_z \theta}}{\sqrt{2 \pi}}
\end{align}

This leads to a 1D-Schr\"odinger equation on $\chi(r)$: 
\begin{align}
 -\frac{\hbar^2}{2 m} \left[ \frac{d^2 \chi}{d r^2} + \frac{1-4 L_z^2}{4 r^2} \chi \right] + \frac{\hbar \omega_c L_z}{2} 
\chi + \frac{m \omega_c^2 r^2}{8} \chi + U_w(r) \chi = E \chi
\end{align}

We introduce dimensionless units: for length $r = x \ell_B = x \sqrt{\frac{\hbar}{m \omega_c}}$ and for energy $E = \epsilon \hbar \omega_c$.
In these units 
\begin{align}
 -\frac{1}{2} \frac{d^2 \chi}{d x^2} + \frac{4 L_z^2-1}{8 x^2} \chi + \frac{L_z}{2} \chi + \frac{x^2}{8} \chi + U_w(x) \chi = \epsilon  \chi
\label{eq:Sh1D}
\end{align}

We want to express this equation as a function of the quasi-classical parameters, we thus use the relation:
\begin{align}
p_\theta &= m \omega_c \frac{R_L^2 - L_c^2}{2} = \hbar L_z \label{eq:S}\\
L_z &= n_L (1 - S^2) 
\end{align}
where we used $R_L= \ell_B \sqrt{2 n_L}$ and introduced the parameter:
\begin{align}
S = \frac{L_c}{R_L}
\end{align}

The Schr\"odinger equation then becomes:
\begin{align}
 -\frac{1}{2} \frac{d^2 \chi}{d x^2} + \frac{4 n_L^2 (1 - S^2)^2-1}{8 x^2} \chi + \frac{n_L(1-S^2)}{2} \chi + \frac{x^2}{8} \chi + U_w(x) \chi = \epsilon  \chi
\end{align}
where $n_L$ is the Landau level and $S = L_c/R_L$. The minimum  $X_m$ of the effective potential energy in Eq.~(\ref{eq:Sh1D}) reads $X_m^4  =  4 L_z^2 - 1$.

Using Eq.~(\ref{eq:S}) we find that collisions with the impurity (in a semiclassical approximation) occur only for orbital momenta $l_z$ in the range:
\begin{align}
l_{min} &= \frac{2 n_L - (\sqrt{2 n_L} + r_d \ell_B^{-1})^2}{2} \\
l_{max} &= \frac{2 n_L - (\sqrt{2 n_L} - r_d \ell_B^{-1})^2}{2}
\end{align}
where $n_L = \epsilon_F/\hbar \omega_c$, this result is useful to estimate the range of orbital momenta that have to be included in the quantum calculation.

\subsection{Schr\"odinger equation in the rotating frame}

The time dependent potential for a linearly polarized field reads:
\begin{align}
V_{ac} = -F_{ac} r \cos \theta \cos \omega t 
\end{align}

In a circularly polarized field we find:
\begin{align}
V_{ac} = -F_{ac} r \cos\left( \theta - \omega t \right) = -F_{ac} r \cos \phi 
\end{align}
where $\phi = \theta - \omega t$, since the potential then depends only on $\phi$ we first concentrate on the circularly polarized case.

We now seek solutions of the Schr\"odinger equation in the rotating wave form:
\begin{align}
\psi = \psi(r,\phi) e^{-i E t/\hbar} 
\end{align}

The Schr\"odinger equation reads:
\begin{align}
i \hbar \frac{d \psi}{d t} &= -i \hbar \omega \frac{\partial \psi}{\partial \phi} + E \psi = {\hat H} \psi \\
{\hat H} \psi &= -\frac{\hbar^2}{2 m} \left[ \frac{1}{r} \frac{\partial}{\partial r} \left( r \frac{\partial \psi}{\partial r} \right) + \frac{1}{r^2} \frac{\partial^2 \psi}{\partial \phi^2} \right] - \frac{i \hbar \omega_c}{2} \frac{\partial \psi}{\partial \phi} + \frac{m \omega_c^2 r^2}{8} \psi + U_w(r) \psi - (F_{ac} r \cos \phi) \psi 
\end{align}
which has the form of a stationary Schr\"odinger equation.

We seek the solution in the form:
\begin{align}
\psi = \sum_{n,L_z} a_{n,L_z} \frac{\chi_{n,L_z}(r)}{\sqrt{r}} \frac{e^{i L_z \phi}}{\sqrt{2 \pi}}  e^{-i E t/\hbar} 
\end{align}
where $\chi_{n,L_z}(r)$ are the Eigenfunctions of the stationary part of the Hamiltonian introduced in the first section.

This leads to the Schr\"odinger equation for $a_{n,L_z}$: 
\begin{align}
(\hbar \omega L_z + E) a_{n,L_z} = \epsilon_{n,L_z} a_{n,L_z} - \sum_{n'} \left[ F(n,L_z;n',L_z+1) a_{n',L_z+1} +  F(n,L_z;n',L_z-1) a_{n',L_z-1} \right]
\label{eq:Circ}
\end{align}
where we introduced the coefficients:
\begin{align}
F(n,L_z;n',L_z') = \frac{F}{2} \int dr \;  \chi_{n,L_z}(r) \chi_{n',L_z'}(r)  r 
\end{align}

The transformation to the rotating frame can also be done in the classical Hamiltonian:
\begin{align}
H &= \frac{p_r^2}{2 m} + \frac{p_\theta^2}{2 m r^2} + \frac{\omega_c}{2} p_\theta + \frac{m \omega_c^2 r^2}{8}  + U_w(r) - F_{ac} r \cos(\theta - \omega t)
\end{align}

We introduce the generating function for the canonical transformation:
\begin{align}
\Phi(r, P_r, \theta, P_\phi; t) = r P_r + (\theta - \omega t) P_\phi 
\end{align}
Then:
\begin{align}
P_r &= p_r \;,\; \phi = \frac{\partial \Phi}{\partial P_\phi} = \theta - \omega t \;,\; p_\theta = \frac{\partial \Phi}{\partial \theta} = P_\phi  \\
H' &= H + \frac{\partial \Phi}{\partial t}\\
&= \frac{P_r^2}{2 m} + \frac{P_\phi^2}{2 m r^2} + \frac{\omega_c}{2} P_\phi + \frac{m \omega_c^2 r^2}{8}  + U_w(r) - F_{ac} r \cos \phi - \omega P_\phi \label{Hcanonic}
\end{align}
which is consistent with the result form quantum mechanics.

\subsection{Quantum master equation}

We now consider the following master equation : 
\begin{align}
\frac{\partial {\hat \rho}}{\partial t} = -\frac{i}{\hbar} [ {\hat H}(t), {\hat \rho} ] - \frac{ {\hat \rho} - {\hat \rho}_{eq} }{\tau} 
\end{align}

The solution of the Schr\"odinger equation takes the form:
\begin{align}
|\psi(t)> &= \psi(r, \theta - \omega t) e^{-i \epsilon t/\hbar} \\
&= |u(t)>  e^{-i \epsilon t/\hbar} 
\end{align}
where in the last line we recognize $ \psi(r, \theta - \omega t)$ as the time periodic  Floquet wave-function $|u(t)>$ and $\epsilon$ is the associated quasi-energy.

We now write the master equation in the basis of the Floquet functions 
\begin{align}
<u_n| \frac{\partial {\hat \rho}}{\partial t} |u_m> &= -\frac{i}{\hbar} \left( \epsilon_n <u_n(t)| - i \hbar \frac{\partial}{\partial t} <u_n(t)| \right) { \hat \rho } |u_m>  \\ \nonumber
&+ \frac{i}{\hbar} <u_n(t)| {\hat \rho} \left(\epsilon_m |u_m(t)> + i \hbar \frac{\partial}{\partial t} |u_m(t)>   \right)  - <u_n| \frac{ {\hat \rho} - {\hat \rho}_{eq} }{\tau} |u_m>\\
 \frac{\partial}{\partial t} \left( <u_n| {\hat \rho}|u_m> \right) &= -i (\epsilon_{n} - \epsilon_m) <u_n| {\hat \rho}|u_m>  - <u_n| \frac{ {\hat \rho} - {\hat \rho}_{eq} }{\tau} |u_m>
\end{align}

where we used:
\begin{align}
{\hat H}(t) |u_n(t)> &=  i \hbar e^{i \epsilon_n t/\hbar} \frac{\partial}{\partial t} \left(  |u_n(t)>  e^{-i \epsilon_n t/\hbar}  \right) \\
 &= \left( \epsilon_n |u_n(t)> + i \hbar \frac{\partial}{\partial t} |u_n(t)>  \right) 
\end{align}

Introducing more compact notations $\rho_{nm} =  <u_n| {\hat \rho}|u_m> $, $\rho_{eq,nm}(t) = <u_n| {\hat \rho}_{eq}|u_m>$ and $\epsilon_{nm} = \epsilon_n - \epsilon_m$, the master equation becomes:
\begin{align}
\frac{\partial}{\partial t} \rho_{nm} + \frac{i }{\hbar} \epsilon_{nm} \rho_{nm} = - \frac{\rho_{nm} - \rho_{eq,nm}(t)}{\tau}
\end{align}

Writing the expression for $\rho_{eq,nm}(t)$ : 
\begin{align}
\rho_{eq,nm}(t) &= \sum_{\epsilon_0(i,L_z) < \epsilon_F} \left( \int (dr d \theta \; r) u_n(r, \theta-\omega t)^* \frac{\chi_{i,L_z}(r)}{\sqrt{r}} \frac{e^{i L_z \theta}}{\sqrt{2 \pi}} \right) \\ \nonumber
&\;\times\; \left( \int (dr' d \theta' \; r') u_m(r', \theta'-\omega t) \frac{\chi_{i,L_z}(r')^*}{\sqrt{r'}} \frac{e^{-i L_z \theta'}}{\sqrt{2 \pi}} \right)  \\
&= \sum_{\epsilon_0(i,L_z) < \epsilon_F} a_{i,L_z}(n)^* a_{i,L_z}(m)
\end{align}
where $a_{i,L_z}(n)$ are the projections of the Floquet wave function on the eigenstates of the unperturbed Hamiltonian which were introduced in the previous section, their expression is:
\begin{align}
a_{i,L_z}(m) = \int (dr' d \theta' \; r') u_m(r', \theta'-\omega t) \frac{\chi_{i,L_z}(r')^*}{\sqrt{r'}} \frac{e^{-i L_z \theta'}}{\sqrt{2 \pi}}
\end{align}
we find that $\rho_{eq,nm}(t)$ is time independent, this is a consequence of the stationary nature of the problem in the rotating frame and of the isotropic character of the equilibrium density matrix.

We thus find that in the rotating frame the density matrix converges to:
\begin{align}
\rho_{nm} = \frac{ \rho_{eq,nm} }{ 1 + \frac{i}{\hbar} \epsilon_{nm} \tau  }
\label{eq:rho}
\end{align}

The charge density distribution in the rotating frame can be expressed through the density matrix 
\begin{align}
n_e(r, \theta_{\cal R}) = \sum_{m n} u_m^*(r, \theta_{\cal R}) \rho_{mn} u_n(r, \theta_{\cal R})
\label{eq:drho}
\end{align}

\subsection{Angle(time) averaged density distribution}

To simplify notations we note $n_e(r) = < n_e(r, \theta) >_\theta$ (where we have omitted the index in $\theta_{\cal R}$).

We are interested in the mean electron density averaged over angles/time 
\begin{align}
n_e(r) &= \sum_{nm} \rho_{nm} <u_n(r, \theta) u_m(r, \theta)^*>_{\theta}
\end{align}

For convenience we introduce the notation:
\begin{align}
U_{mn}(r) =  <u_m(r, \theta)^* u_n(r, \theta)>_{\theta} 
\end{align}

Using the decomposition of the Floquet waves in the stationary eigenbasis we find:
\begin{align}
U_{mn}(r) &= < \sum_{i,L_z} a_{i,L_z}(n) \frac{\chi_{i,L_z}(r)}{\sqrt{r}} \frac{e^{i L_z \theta}}{\sqrt{2 \pi}} \sum_{j,L_z'} a_{j,L_z'}(m)^* \frac{\chi_{j,L_z'}(r)^*}{\sqrt{r}} \frac{e^{-i L_z' \theta}}{\sqrt{2 \pi}} >_\theta \\
&= \frac{1}{2\pi} \sum_{i,j;L_z} a_{j,L_z}(m)^* a_{i,L_z}(n)  \frac{\chi_{j,L_z}(r)^* \chi_{i,L_z}(r) }{r} 
\end{align}

With the notation 
\begin{align}
n_e(r)=  \sum_{nm} \rho_{nm} U_{mn}(r) 
\end{align}
and using Eq.~(\ref{eq:rho}) we find a closed form expression for the angle/time averaged density: 
\begin{align}
n_e(r) = \sum_{nm} \frac{U_{mn}(r)}{ 1 + \frac{i}{\hbar} \epsilon_{nm} \tau  } \sum_{\epsilon_0(i,L_z) < \epsilon_F} a_{i,L_z}(n)^* a_{i,L_z}(m)
\end{align}

For a consistency check, we set $\tau = 0$ then:
\begin{align}
n_e(r) &= \sum_{\epsilon_0(i,L_z) < \epsilon_F}  \sum_{nm}  a_{i,L_z}(m) U_{mn}(r) a_{i,L_z}(n)^* \\
 &= \sum_{\epsilon_0(i,L_z) < \epsilon_F}  \frac{1}{2\pi}  \frac{|\chi_{i,L_z}(r)|^2}{r} 
\end{align}
due to orthogonality relations $\sum_{n}  a_{i,L_z}(n)^* a_{j,L_z'}(n) = \delta_{ij} \delta_{L_z,L_z'}$. 

Another check is that we must recover the equilibrium density if there is no microwave, in this case 
$a_{i,L_z}(n) = \delta_{i,L_z;n}$ which leads to: 
\begin{align}
&\sum_{\epsilon_0(i,L_z) < \epsilon_F} a_{i,L_z}(n)^* a_{i,L_z}(m) = \delta_{nm} \sum_{\epsilon_0(i,L_z) < \epsilon_F} \delta_{i,L_z;n}\\
&U_{nn}(r) = \frac{1}{2\pi} \sum_{i,L_z} \delta_{i,L_z;n} \frac{|\chi_{i,L_z}(r)|^2 }{r} 
\end{align}
We then have :
\begin{align}
n_e(r) &= \sum_{n}  U_{nn}(r) \sum_{\epsilon_0(i,L_z) < \epsilon_F} \delta_{i,L_z;n}\\
&= \sum_{\epsilon_0(i,L_z) < \epsilon_F}  \frac{1}{2\pi}  \frac{|\chi_{i,L_z}(r)|^2}{r} 
\end{align}

The previous form is not so convenient for the efficient calculation of 
the dependence on $r$ :
\begin{align}
n_e(r) &= \frac{1}{2 \pi r} \sum_{nm}  \frac{1}{ 1 + \frac{i}{\hbar} \epsilon_{nm} \tau  } \sum_{\epsilon_0(i',L_z') < \epsilon_F} a_{i',L_z'}(n)^* a_{i',L_z'}(m) 
\sum_{i,j;L_z} a_{j,L_z}(m)^* a_{i,L_z}(n)  \chi_{j,L_z}(r)^* \chi_{i,L_z}(r) \\
n_e(r) &= \frac{1}{2 \pi r} \sum_{i,j,L_z} {\cal N}_{i,j,L_z}\chi_{j,L_z}(r)^* \chi_{i,L_z}(r)
\end{align}
where 
\begin{align}
{\cal N}_{i,j,L_z} =  \sum_{nm} \frac{a_{j,L_z}(m)^* a_{i,L_z}(n)}{ 1 + \frac{i}{\hbar} \epsilon_{nm} \tau } \sum_{\epsilon_0(i',L_z') < \epsilon_F} a_{i',L_z'}(n)^* a_{i',L_z'}(m) 
\end{align}

\subsection{Typical numerical parameters}

We provide a complete list of the numerical parameters for the charge density vortex calculation in Fig.~4.a from the main article, physical parameters are $J = 2.7$, $r_d/\ell_B=2$, $U_{disc} = 90 \hbar \omega_c$, $qE_{ac}\ell_B/\hbar \omega_c = 0.1$ and $\epsilon_F/\hbar \omega_c=40$ with a relaxation time $\omega_c \tau=10$.

We select a finite set of Landau levels $n \le N_L$ ($N_L = 100$) and of orbital momenta $L_{min} \le l_z \le L_{max}$ with $L_{min} =  -550$ and $L_{max} = 650$, 
for all $l_z$ in this range the one dimensional Schr\"odinger Eq.~(\ref{eq:Sh1D}) is solved using a discretisation with a space step (in dimensionless units) $0.02$ with a total of $8000$ steps, this allows to find the $N_L$ lowest eigenvalues $\epsilon^{(0)}_{n, l_z}$ and eigenvectors $\chi_{n, l_z}$  for each $l_z$. We then filter the states depending on their distance to the Fermi level selecting only states with $| \epsilon^{(0)}_{n, l_z} - \epsilon_F | <= 8 \hbar \omega_c$, this gives a basis with a total of $9400$ states. The Schr\"odinger Eq.~(\ref{eq:Circ}) is then expressed in this basis and diagonalised using full diagonalisation routines from the eigen++ packages. This allows us to find the density matrix using Eq.~(\ref{eq:rho}) and the charge distribution in the rotating frame using Eq.~(\ref{eq:drho}). We have checked that the numerical results are not changed when the basis is expanded to include more states in the calculation, another independent check is the good agreement we find with semiclassical calculations. 

\section*{2. Husimi distribution}
\label{sisec2}

The Husimi function is defined via 
\begin{align}
 H(l_z, \theta) = \sum_{n}  \int_{-\pi}^{\pi} d\theta' e^{-( \theta - \theta' - 2\pi n)^2/2} e^{i l_z \theta'} \psi_{r_d}(\theta') 
\label{eq:s1}
\end{align}
where $\psi_{r_d}(\theta')$ describes the orbital harmonics of the wavefunction near the disc (with radius $r_d$). 

For a hard disc potential the wavefunction $\psi(r, \theta)$ tends to vanish for $r \rightarrow r_d$, we have thus followed 
the approach of \cite{Husimi1S,Husimi2S} where $\psi_{r_d}$ is given (up to normalization) by:
\begin{align}
\psi_{r_d}(\theta) = \partial_r \psi(r = r_d, \theta)
\label{eq:s2}
\end{align}

This expression does not separate incoming and outcoming plane waves incident on the disc, an approximate separation can be achieved by defining $\psi_{r_d}$ as: 
\begin{align}
\psi_{r_d}(\theta) = \sum_{n_l,l_z} a_{n_l,l_z} \frac{1}{k_{l_z,n_l}}\frac{d R_{l_Z,n_l}(r = r_d)}{d r} e^{i l_z \theta}
\label{eq:s3}
\end{align}
where the coefficients $a_{n_l,l_z}$ give the components of the wavefunction in the eigenbasis $R_{l_Z,n_l}(r)e^{i l_z \theta}$ of the system without microwave field:
\begin{align}
\psi(r, \theta) = \sum_{n_l,l_z} a_{n_l,l_z} R_{l_Z,n_l}(r) e^{i l_z \theta}
\label{eq:s4}
\end{align}
and gives the semiclassical momentum at the collision 
\begin{align}
k_{l_z,n_l} = \sqrt{2(E_{l_z,n_l} - U(l_z, r_d))/m} 
\label{eq:s15}
\end{align}
with $E_{l_z,n_l}$ the laboratory frame energy (without microwaves) corresponding to eigenfunction $R_{l_Z,n_l}(r)e^{i l_z \theta}$.

We have checked that the two approaches give similar results in the limit $r_d \ll r_c$. 
In Fig.~1(a) we use the representation (\ref{eq:s3}) which
capture the separate contribution of the in-going wave. This can be seen from the 
representation of two in-going/out-going waves $\psi(r) \propto A_k (\exp(i k (r-r_d)) -  \exp(-i k (r-r_d)))$
with a certain amplitude $A_k$
and the condition $\psi(r=r_d)=0$. 

\section*{3. Solution of the kinetic equations}

The kinetic equation in the laboratory frame reads: 
\begin{align}
& \frac{\partial f}{\partial t} + \mathbf{v} \cdot \frac{\partial f}{\partial \mathbf{r}} + \left[ {\mathbf \omega}_c \times  \mathbf{v} - \frac{1}{m} \partial_r U_w + \frac{1}{m} \mathbf{F}(t) \right]  \cdot \frac{\partial f}{\partial \mathbf{v}} = - \frac{f - f_0}{\tau} \label{eq:kin} \\
& \frac{\partial f}{\partial t} + \left\{f, H(t)\right\} = \frac{d f}{d t} = - \frac{f - f_{eq}}{\tau} \label{eq:poisson}
\end{align}
where $d/dt$ is the derivative along a trajectory in phase space. 

By moving to the rotating frame, the kinetic equation becomes stationary in the same way as the quantum master equation, this can be seen for example by making the canonic transform Eq.~(\ref{Hcanonic}) on Eq.~(\ref{eq:poisson}). The distribution function $f$ thus converges to a steady state value $f({\tilde r}, {\tilde v})$ in the rotating frame coordinates ${\tilde r} = r e^{-i \omega t}$ and  ${\tilde v} = v e^{-i \omega t}$ ($r$ and $v$ are the position coordinates/velocity in the laboratory frame in the complex notation $r = x + i y$ and $v = v_x + i v_y$). We find the distribution function by integrating the equation $\frac{d f}{dt} = - \frac{f - f_0}{\tau} $ along trajectories in the rotating frame. 

Trajectories outside the impurity are found by exact integration of the free equations motion in  magnetic and microwave fields, they give the following expression for the change of charge velocity from ${\tilde v}_i$ at time $t_i$ to $ {\tilde v}_f$
at time $t_f$ after a time interval $\Delta t = t_f - t_i$.
\begin{align}
{\tilde v}_f &= \frac{i F_{ac}}{\omega_c - \omega} + \left({\tilde v}_i - \frac{i F_{ac}}{\omega_c - \omega} \right) e^{i (\omega_c - \omega) \Delta t}\\
{\tilde r}_f &= {\tilde r}_i e^{-i \omega \Delta t} + 
\frac{i {\tilde v}_i}{\omega_c} e^{-i \omega \Delta t} \left( 1 -  e^{i \omega_c \Delta t}  \right) + F_{ac} 
\left( \frac{e^{-i (\omega-\omega_c) \Delta t}}{\omega_c (\omega - \omega_c)}
 - \frac{e^{-i \omega \Delta t}}{\omega \omega_c} 
 - \frac{1}{\omega(\omega-\omega_c)}
 \right) 
\label{eq:step}
\end{align}
The time step in the stepping algorithm is adapted as function of the distance to the impurity and collision times are found with a Newton method. For repulsive impurities a specular reflection occur at the interface, for an attractive potential the particle is propagated inside the impurity using Eq.~(\ref{eq:step}) hold inside the impurity for a box potential.

This procedure allows us to find the distribution function for energies close to the Fermi energy on a four dimensional grid of size $N_r \times N_\theta \times N_\chi \times N_E$ where $N_r = 160$ is the number of entries for the discretization along the polar distance $r$, $N_\theta = 100$ the number of entries for the polar angle $\theta$, $N_\chi = 30$ the number of entries for the angle $\chi$ between position and velocity vectors and $N_E = 30$ the number of entries for energy dicretization centered around the Fermi energy (parameters are given for calculation of Fig.~4b in the main text).

\section*{4. Derivation of the collision map}
\label{sisec3}

The geometry of elastic collision with a hard disc is shown in Fig.~1(a)
and Fig.~\ref{figS1}. The collision angles shown in these Figs. are defined as

\begin{align}
\left\{
\begin{array}{cc}
\theta_c &= 2 \alpha - \theta - 2 \beta + \pi\\
\alpha_c &= \alpha + 2\pi - 2 \beta 
\end{array}
\right.
\end{align}
where 
\begin{align}
\beta = \arg\left(1 + i\frac{r_d}{r_c} e^{-i (\theta - \alpha)} \right)
\end{align}

\begin{figure}
\includegraphics[width=0.50\columnwidth]{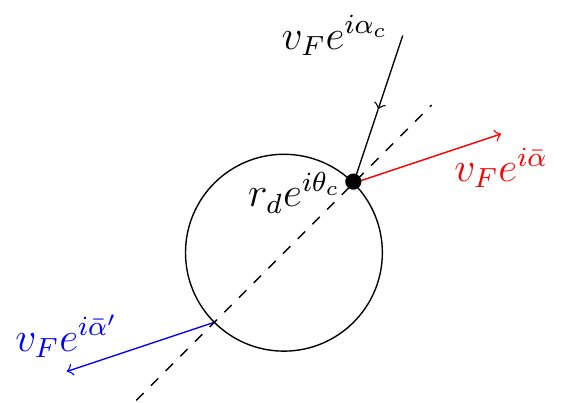}
\caption{Geometry of reflection during collision with a disc in case of attractive and repulsive potentials.
}
\label{figS1}
\end{figure}

After the collision $r = |r| e^{i \theta}$ , $v = |v| e^{i \alpha}$ , $\chi = \alpha - \theta$. 
Without microwaves: conservation of angular momentum and energy implies that $\chi$ is conserved during a collision: ${\bar \chi} = \chi$. 
New values for $\alpha$ can be found from geometrical arguments 
\begin{align}
{\bar \alpha} &= \alpha + \sigma(l_z) \\
l_z &= r_d v_F \sin \chi \\
\sigma &= 2 \chi - 2 \arctan\left(  \frac{\tan \chi}{\sqrt{1 - \frac{2 U}{m v_F^2 \cos^2 \chi}}} \right)
\end{align}
with $U$ the amplitude of the impurity potential ($\sigma, \chi \in (-\pi/2, \pi/2)$).

In the special case of a collision with a hard disc repulsive potential ($U \rightarrow \infty$) :
\begin{align}
\sigma &= 2 \chi - \pi
\end{align}

for an attractive disc potential with a large amplitude  (amplitude $U \rightarrow -\infty$) this gives instead.
\begin{align}
\sigma &= 2 \chi 
\end{align}

The change in kinetic energy during a period due to microwaves is :
\begin{align}
{\bar E} - E = \int_0^{2 \pi/\omega_c} F_{ac} {\rm Re}\left(v_F e^{i \alpha_R} e^{i (\omega_c - \omega) t} \right) dt  = v_F F_{ac} \frac{\sin \alpha_R - \sin(\alpha_R - 2 \pi J)}{ \omega - \omega_c }
\end{align}
however 
\begin{align}
E \simeq H + \omega r_d \sqrt{2 m H} \sin \chi 
\end{align}
where $H \simeq \epsilon_F$ is conserved during the time evolution we thus find 
\begin{align}
\sin {\bar \chi} - \sin \chi =  \frac{F_{ac}}{r_d \omega (\omega - \omega_c)} \left[ \sin \alpha_R - \sin(\alpha_R - 2 \pi J) \right]
\end{align}

This leads to the following map:
\begin{align}
\left\{
\begin{array}{cl}
\sin {\bar \chi} &= \sin \chi +  \frac{F_{ac}}{r_d \omega (\omega - \omega_c)} \left[ \sin \alpha_R - \sin(\alpha_R - 2 \pi J) \right]\\
{\bar \alpha}_R &= \alpha_R + 2 {\bar \chi} - 2 \sigma({\bar \chi}) - 2 \pi J
\end{array}
\right. \label{eq:stdmap}
\end{align}
Under the approximation $\sin \chi \simeq \chi$, $\sigma(\chi) = \pi/2$ (hard disc) we recover the previous (approximate map) for $J > 0$ (check $J < 0$)

We now determine the kinetic energy from the standard map using 
the energy conservation in the rotating frame:
\begin{align}
H = \frac{m v^2}{2} - F_{ac} R_d \cos \phi - \omega P_\phi 
\end{align}
where $m v^2/ 2$ is the kinetic energy, $\phi$ the angle in the 
rotating frame and $P_\phi$ its conjugate momentum.

We need to express the rotating frame variables from the standard 
map variables, we find : 
\begin{align}
\phi &= \alpha_R - \chi\\
P_\phi &=  m v R_d \sin \chi - \frac{m R_d^2 \omega_c}{2} 
\end{align}

We thus find an equation on $v$: 
\begin{align}
H = \frac{m v^2}{2} - (m \omega R_d) v \sin \chi - F_{ac} R_d \cos \phi + \frac{m R_d^2 \omega \omega_c}{2} 
\end{align}

To zero order in $R_d$, $E = m v^2/2 = H$, going to first order in $R_d$ we find:
\begin{align}
E = H +  \omega R_d \sqrt{2 m H} \sin \chi + F_{ac} R_d \cos \phi 
\end{align}
at the Fermi energy $F_{ac}/(m \omega v_F) \ll 1$ so we can approximate: 
\begin{align}
E \simeq H +  \omega R_d \sqrt{2 m H} \sin \chi 
\end{align}
The exact solution is:
\begin{align}
v = m \omega R_d  \sin \chi + \sqrt{ 2 H + 2 F_{ac} R_d  \cos(\alpha_R - \chi) + m^2 R_d^2 \omega^2  \sin^2 \chi - R_d^2 \omega \omega_c } 
\label{eq:vmap}
\end{align}

We can also determine the trajectory parameters $R_L, L_c$ : 
\begin{align}
R_L &= \frac{1}{|\omega_c|} \sqrt{ \frac{P_r^2}{m^2} + \left( \frac{P_\phi}{m r} + \frac{\omega_c r}{2} \right)^2 } \\
L_c &= \frac{1}{|\omega_c|} \sqrt{ \frac{P_r^2}{m^2} +  \left( \frac{P_\phi}{m r} - \frac{\omega_c r}{2} \right)^2  } 
\end{align}
where $K$ is the kinetic energy.


\section*{5. Calculations of absorbed microwave power}
\label{sisec4}

We first provide the derivation of the approximate relation between absorbed microwave power and the rotating dipole moment around the impurity.
\begin{align}
{\cal P} &= {\rm Tr} ({\hat \rho} \mathbf{v}\cdot q\mathbf{E}_{ac} ) \\
&= -\frac{i}{\hbar} {\rm Tr} ({\hat \rho} [ \mathbf{r}, {\hat H}_0 ] \cdot q\mathbf{E}_{ac} ) \\
&= -\frac{i}{\hbar} {\rm Tr} ({\hat \rho} [ \mathbf{r}, {\hat H}_{\cal R} 
+ \omega {\hat l}_z  ] \cdot q\mathbf{E}_{ac} ) \\
&= -\frac{i}{\hbar} {\rm Tr} ({\hat \rho} [ \mathbf{r}, \omega {\hat l}_z  ] \cdot q\mathbf{E}_{ac} ) 
+ \frac{i}{\hbar} {\rm Tr} ( \mathbf{r} [{\hat \rho}, {\hat H}_{\cal R}  ] \cdot q\mathbf{E}_{ac} ) \\
&= \omega {\rm Tr} ({\hat \rho} \mathbf{r})  \cdot ( \mathbf{e}_z \times q \mathbf{E}_{ac} )  + O(1/\tau) \; .
\label{eq:powerSI4}
\end{align}

In numerical quantum master equation calculations the power was estimated using:
\begin{align}
{\cal P} &= -q\mathbf{E}_{ac} \cdot \frac{i}{\hbar} {\rm Tr} ({\hat \rho} [ \mathbf{r}, {\hat H}_0 ] ) 
\label{eq:powernum}
\end{align}
for high values of $\omega_c \tau$ we found a good agreement between exact calculations using Eq.~\ref{eq:powernum} and estimations from the dipole moment Eq.~\ref{eq:powerSI4}. 

For the semiclassical kinetic equation we can compute the average absorbed power using the relation:
\begin{align}
{\cal P} &= \left<\int d^2\mathbf{v} \int d^2\mathbf{r} \left[ H_0(\mathbf{v}, \mathbf{r}) - E_F \right] \frac{f(\mathbf{r}, \mathbf{v}, t) - f_0(\mathbf{r}, \mathbf{v})}{\tau}\right>_t \label{eqPowsemi} \\ 
&= \left<\int d^2\mathbf{v} \int d^2\mathbf{r} \left[ \frac{m \mathbf{v}^2}{2} + U_w(\mathbf{r}) \right] \frac{f(\mathbf{r}, \mathbf{v}, t) - f_0(\mathbf{r}, \mathbf{v})}{\tau}\right>_t \\
&= -\left<\int d^2\mathbf{v} \int d^2\mathbf{r} \left[ \frac{m \mathbf{v}^2}{2} + U_w(\mathbf{r}) \right] \frac{d f}{d t}\right>_t\\
&=  \left<\int d^2\mathbf{v} \int d^2\mathbf{r} \frac{d}{dt}\left[ \frac{m \mathbf{v}^2}{2} + U_w(\mathbf{r}) \right] f \right>_t\\
&=  \left<\int d^2\mathbf{v} \int d^2\mathbf{r} [ \mathbf{v} \cdot \mathbf{F}(t) ] f \right>_t \label{eqPowdir}
\end{align}
where $H_0$ is the Hamiltonian without microwaves. 

The advantage of equation Eq.~(\ref{eqPowsemi}) is that the function that is integrated is always positive as opposed to Eq.~(\ref{eqPowdir}) which has cancellations between positive and negative terms,  Eq.~(\ref{eqPowsemi}) is thus more suited to Monte-Carlo type of evaluation.

\begin{figure}[h]
\includegraphics[width=0.65\columnwidth]{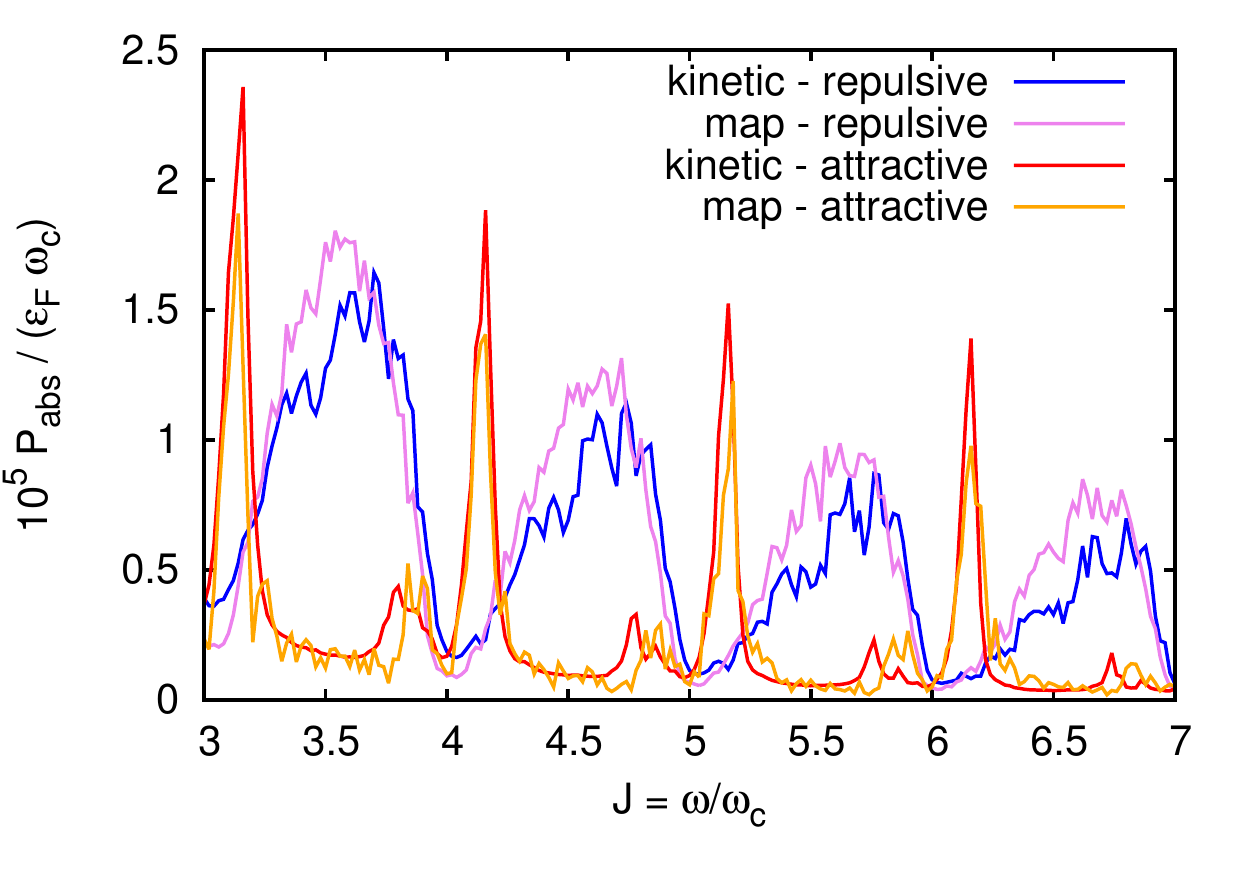}
\caption{Dependence of the average absorbed microwave power $P_{abs}$ for electrons colliding with an impurity for parameters: $q E_{ac}/(m \omega v_F) = 0.013$, $r_d/R_c = 0.13$, $\omega_c \tau = 100$, $U_{disc} = \pm 2 \epsilon_F$, a good agreement is observed between results from the kinetic equation and from the collision map \ref{eq:stdmap}. 
}
\label{figS2}
\end{figure}

Finally it is possible to estimate the absorbed power from the standard map Eq.~(\ref{eq:stdmap}), in this case the exact dynamics over one period is replaced by the approximate map; the distribution function is then computed by integrating $\frac{d f}{d t} = - \frac{f - f_{eq}}{\tau}$ for free evolution with the map parameters during time $2 \pi/\omega_c$ which separates successive collision events (in the approximation $r_d \ll R_c$). The absorbed power is then also computed from Eq.~(\ref{eq:powernum}). 
The results presented in Fig.~\ref{figS2} show that the absorbed power obtained from the kinetic equation is 
in a good agreement with the results based on the approximate collision map  Eq.~(\ref{eq:stdmap}). This shows how physically relevant quantities can be computed from the simplified map dynamics. We note the significant difference between the case of attractive and repulsive impurity potentials.

The dependence  ${\cal P(J)}$ at various sizes
of repulsive impurity potential $r_d$ is shown in Fig.~\ref{figS3}.
These data show a good agreement
between the quantum and classical simulations
until the impurity diameter $2r_d$ remains larger than 
the magnetic length $\ell_B$.
\begin{figure}[h]
\includegraphics[width=0.65\columnwidth]{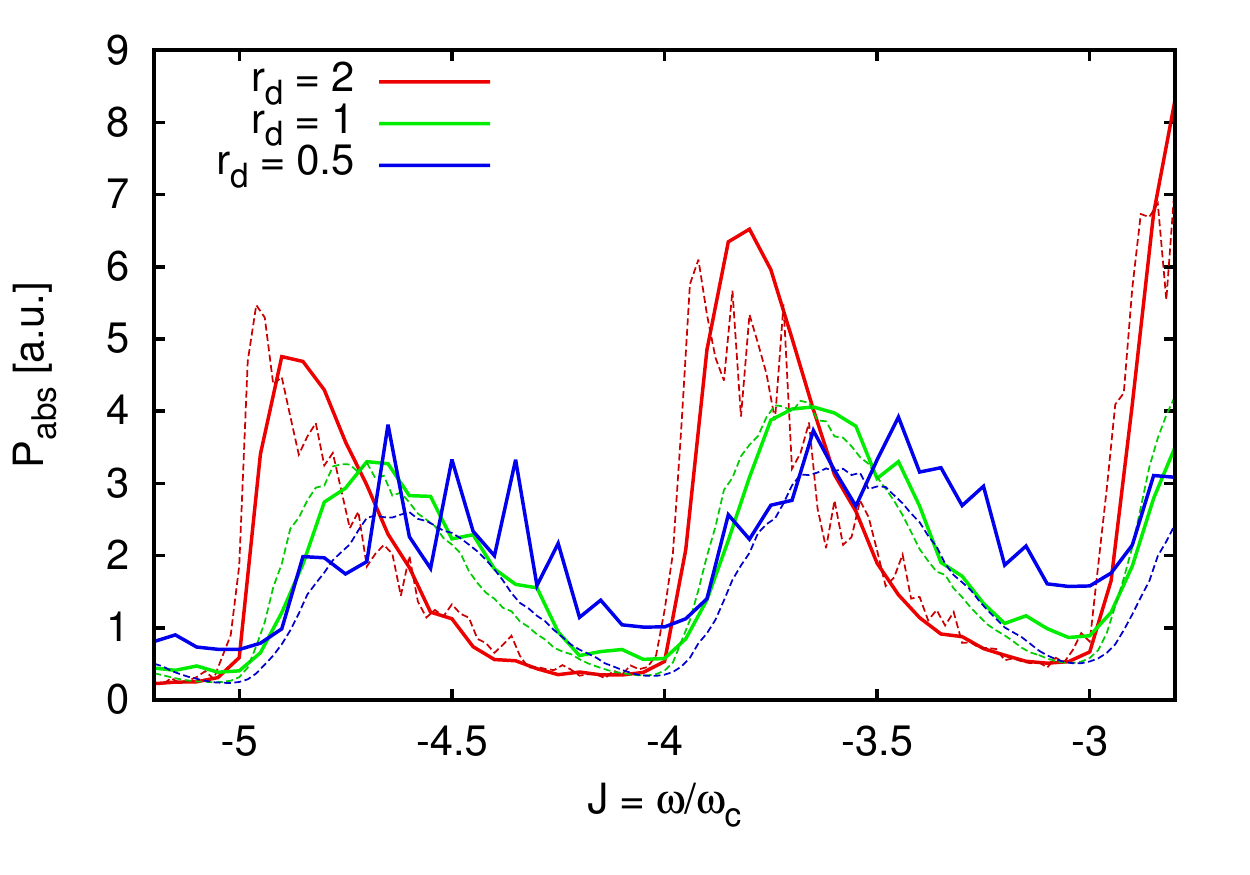}
\caption{Dependence of the microwave absorption  
${\cal P}_{abs}$ on $J$ (arbitrary units) for different values of impurity radius
$r_d$ obtained from the quantum master equation Eq.~(\ref{eq:powernum}) (full curves) and 
the classical kinetic equation Eq.~(\ref{eqPowsemi}) (dashed curves).
Here $U_{disc} = 2 \epsilon_F$, $qE_{ac} \ell_B/\hbar \omega_c =0.1$ and $\omega_c \tau =100$. The semiclassical curves reproduce correctly the results from the quantum master equation. For $r_d/\ell_B = 0.5$ additional fluctuations appear in the quantum calculation around an average lineshape which is still well describe by the semiclassical calculation. We attribute those to interference effects which become more pronounced when the size of the impurity becomes closer to the Fermi wavelength. In a real sample such fluctuations would probably be suppressed by ensemble average leaving only an average average value which would correspond to the semiclassical result. 
}
\label{figS3}
\end{figure}

The results obtained from the quantum master equation (6)
show that at small microwave driving the absorption power scales as
${\cal P} \propto {E_{ac}}^2 $ 
(see Fig.~\ref{figS4}). The results of numerical simulations of 
classical kinetic equation give the same scaling.

\begin{figure}[h]
\includegraphics[width=0.65\columnwidth]{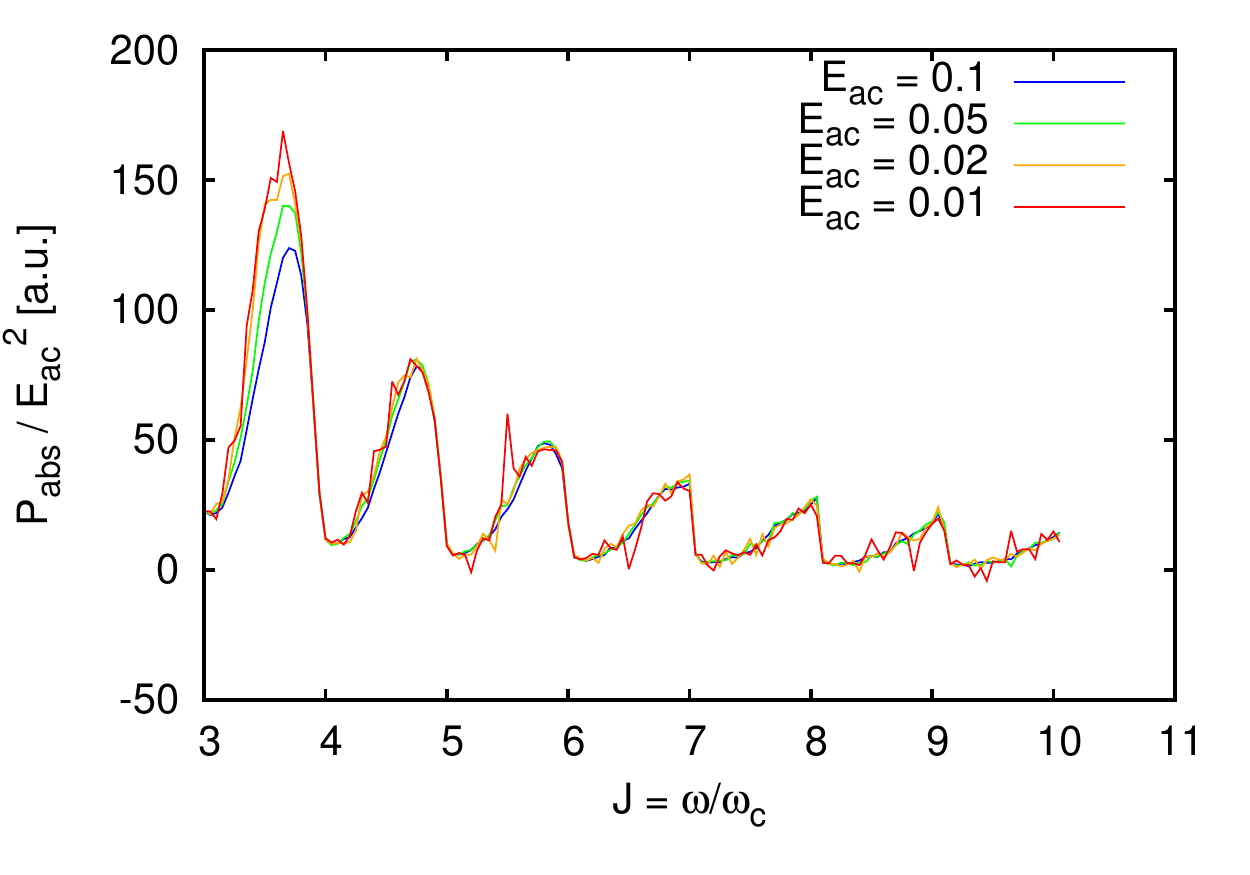}
\caption{Scaling dependence of the microwave absorption  
${\cal P}_{abs}$ rescaled by incident microwave power $\propto E_{ac}^2$ on $J$ (arbitrary units) for different amplitudes of microwave field $E_{ac}$. The results are obtained from the quantum master equation Eq.~(\ref{eq:powernum}) for a repulsive impurity potential ($U_d = 90 \hbar \omega_c$) at $\omega_c \tau =100$ and $0.01 \leq E_{ac} \leq 0.1$. The data show that the scaling $P_{abs} \propto E_{ac}^2$ is highly accurate. At the lowest microwave powers the data become more noisy and start to shows individual resonances, we think that the sharp peaks comes from contributions of individual levels trapped around the impurity that lead to sharp resonances due to the high $\omega_c \tau = 100$ value. At higher microwave power the individual resonances are broadened  giving the semiclassical absorption curve. 
}
\label{figS4}
\end{figure}

\clearpage 
\section*{6. Estimates for charge density variation}
\label{sisec6}

In general we can write the following scaling form for the charge density vortex : 
\begin{align}
\frac{\delta n_e}{n_e} = f_7( x R_c^{-1}, y R_c^{-1}, r_d  R_c^{-1}, \frac{q E_{ac}}{m \omega_c v_F}, \frac{\omega}{\omega_c}, \omega_c \tau, \frac{\epsilon_F}{\hbar \omega_c} )
\end{align}
where $f_7$ is a dimensionless function of its seven dimensionless arguments. 

In the semiclassical limit the parameter $\frac{\epsilon_F}{\hbar \omega_c}$ is not relevant, this parameter describes the amplitude of microwave induced Friedel oscillations. Also in the linear response regime $\frac{\delta n_e}{n_e} \propto  \frac{q E_{ac}}{m \omega_c v_F}$ (due to the connection between the rotating dipole and microwave power absorption this scaling is related to the $\propto E_{ac}^2$ scaling for the absorbed power, its accuracy is shown on Fig.~\ref{figS4}).

This leads to the following scaling form:
\begin{align}
\frac{\delta n_e}{n_e} = \frac{q E_{ac}}{m \omega_c v_F} f_5( x R_c^{-1}, y R_c^{-1}, r_d  R_c^{-1}, \frac{\omega}{\omega_c}, \omega_c \tau )
\end{align}
where $f_5$ is a dimensionless function of the remaining five arguments.

Figs.~\ref{figSRd},\ref{figSJ},\ref{figStau} show the scaling dependence on the remaining parameters $r_d  R_c^{-1}$, $\frac{\omega}{\omega_c}$ and $\omega_c \tau$. 

Fig.~\ref{figSRd} shows that $\delta n_e$ only weakly depends on $r_d R_c^{-1}$, this is perhaps the most surprising result since the number of electrons colliding with a given impurity during a cyclotron period is $\sim n_e R_c r_d$, this would suggest that $\delta n_e \propto n_e R_c r_d$. Our simulations show that at least in the semiclassical limit, this argument does not hold when the map parameter is small $\epsilon = \frac{q E_{ac}}{r_d \omega (\omega - \omega_c)} \ll 1$  (see Eq.~(\ref{eq:stdmap})), in this case it seems that the $\propto r_d^{-1}$ dependence from the kick amplitude in the standard map equations partially cancels the $\propto r_d$ dependence from the collision cross section. Of course in the limit $r_d \rightarrow 0$ the map parameter $\epsilon$ diverges so that this cancellation holds only for sufficiently small excitation field $E_{ac}$. Other figures show an approximate $1/J$ dependence (for a fixed value of ${\rm mod_1 J}$) in Fig.~\ref{figSJ} and a weak dependence on $\omega_c \tau$ in Fig.~\ref{figStau}. Combining these results together we find the following scaling relation:
\begin{align}
\frac{\delta n_e}{n_e} = \frac{q E_{ac}}{m \omega v_F} f_2( x R_c^{-1}, y R_c^{-1} )
\end{align}
which is the result given in the main text.

\begin{figure}[h]
\includegraphics[width=0.65\columnwidth]{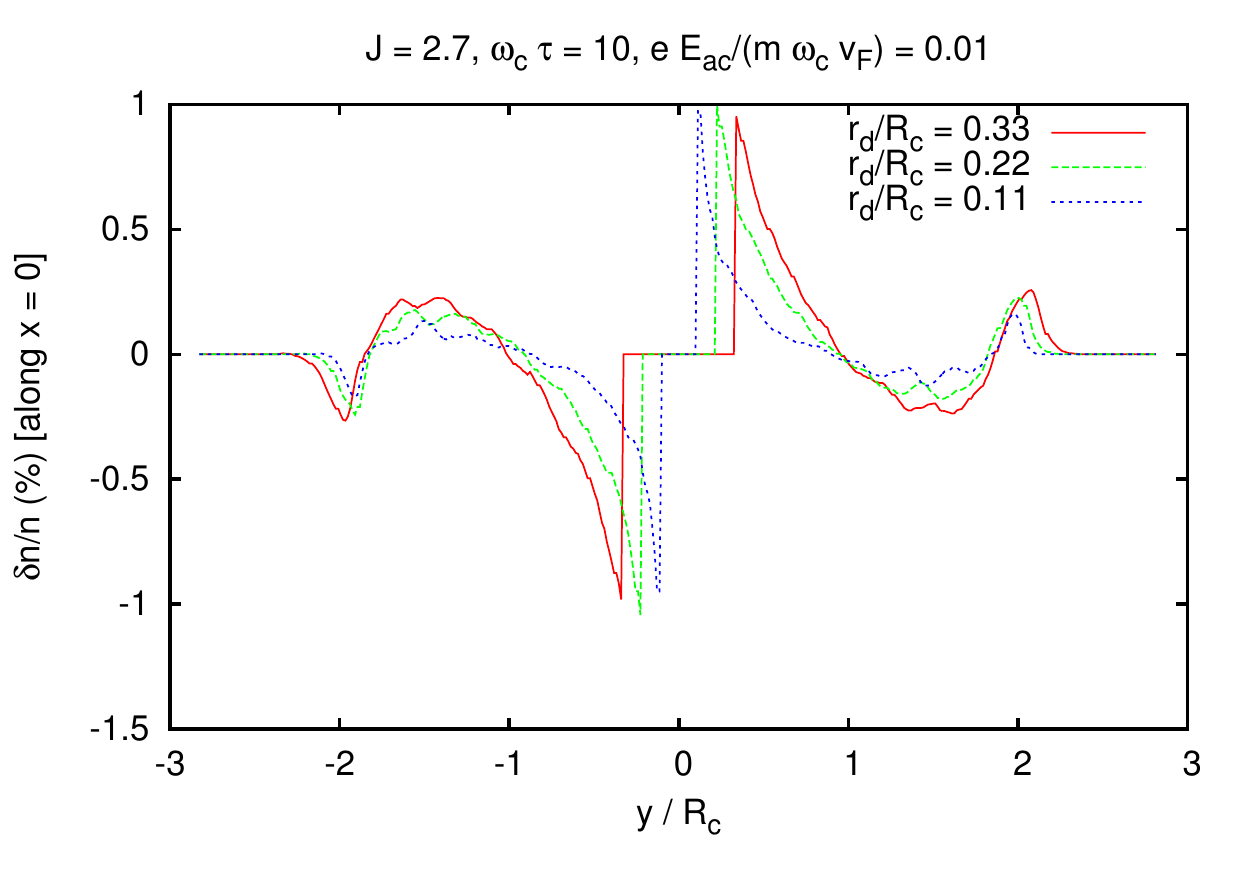}
\caption{Dependence of $\delta n_e(x = 0, y)/n_e$ on $y/R_c$ for different impurity diameters, results are shown for the kinetic equation calculation.
}
\label{figSRd}
\end{figure}

\begin{figure}[h]
\includegraphics[width=0.65\columnwidth]{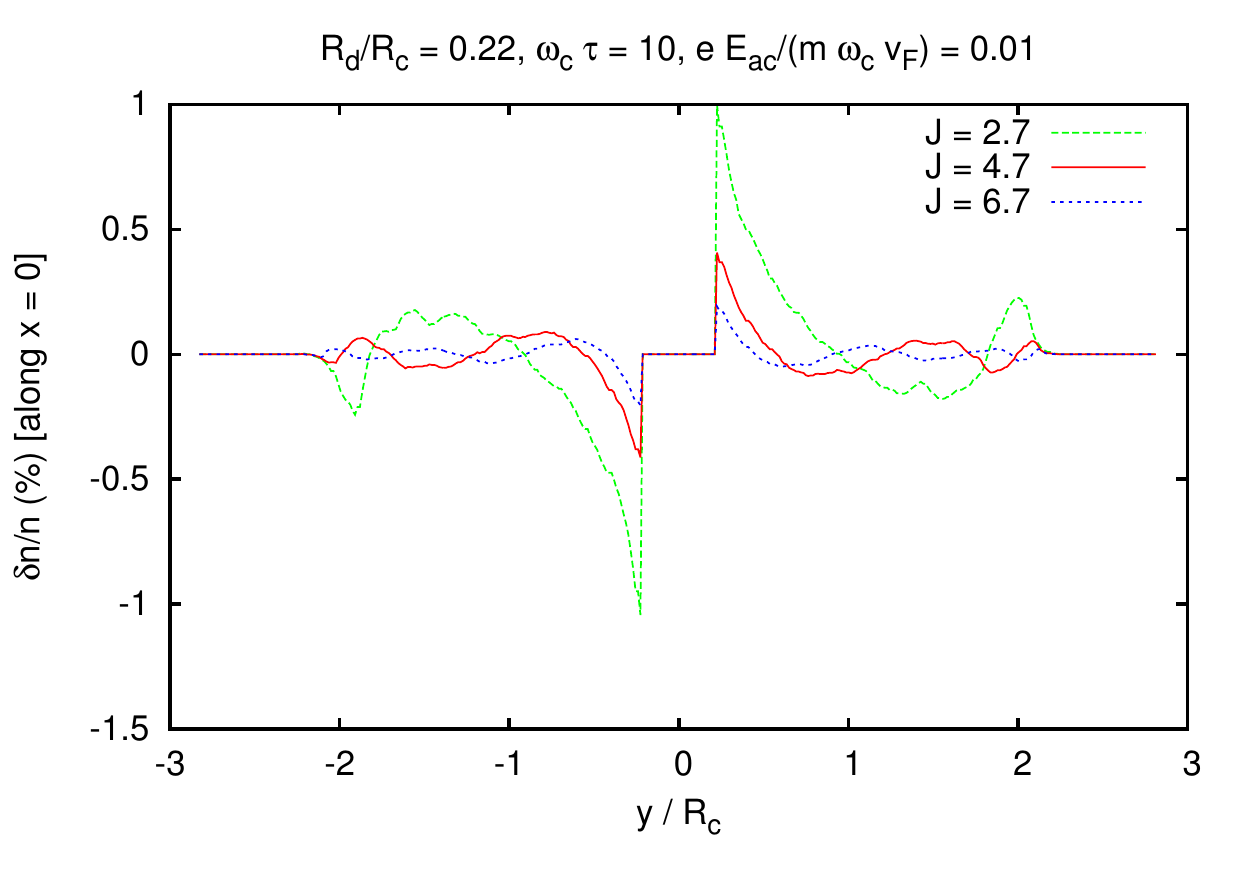}
\caption{Dependence of $\delta n_e(x = 0, y)/n_e$ on $y/R_c$ for different values of $J = \omega/\omega_c$.
}
\label{figSJ}
\end{figure}

\begin{figure}[h]
\includegraphics[width=0.65\columnwidth]{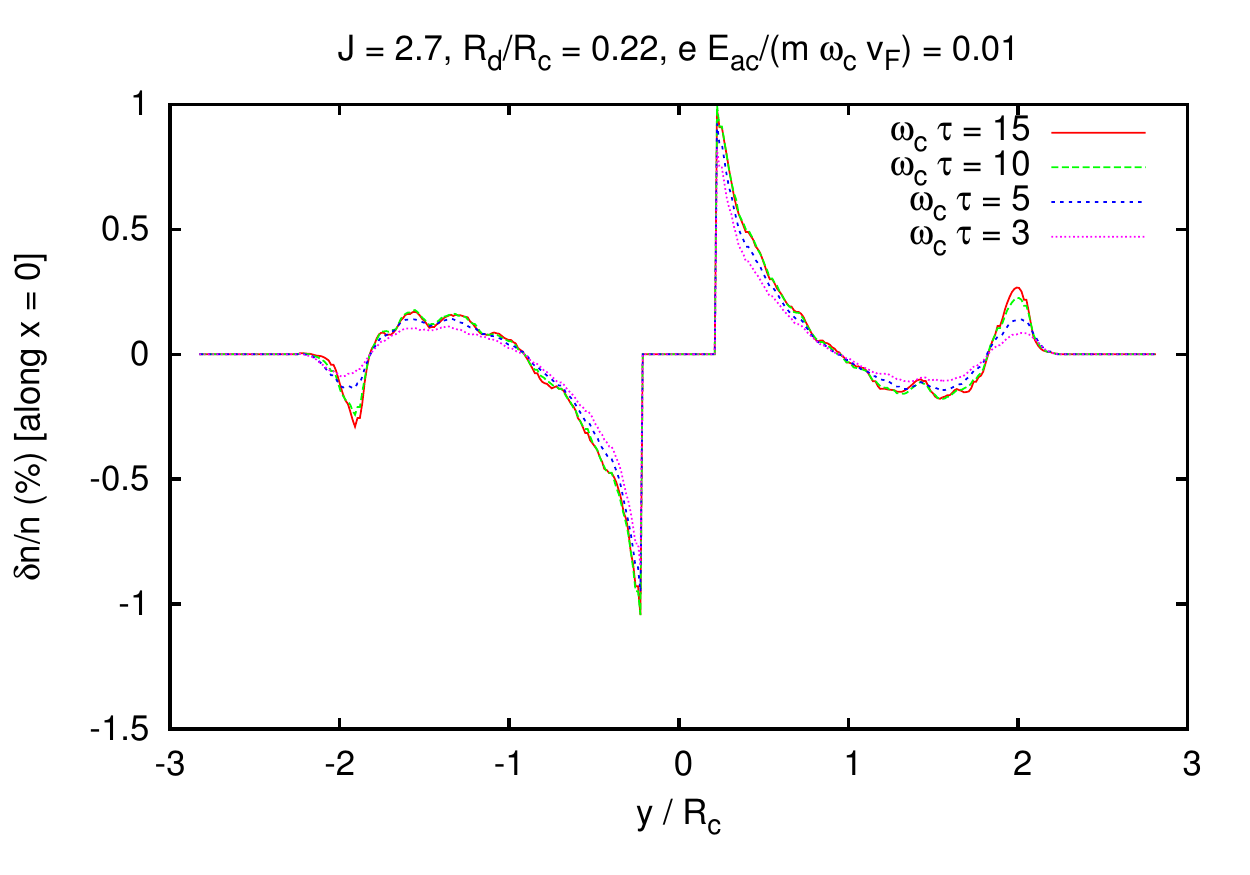}
\caption{Dependence of $\delta n_e(x = 0, y)/n_e$ on $y/R_c$ for different values of $\omega_c \tau$.
}
\label{figStau}
\end{figure}

\section*{References}

\bibitem{sh2d} G. Paz, Eur. J. Phys. {\bf 22}, 337-341 (2001) {\it On the connection between the radial momentum operator and the Hamiltonian in n-dimensions} 

\bibitem{LandauT3} L. Landau and E. Lifshitz, {\it Quantum mechanics : non-relativistic theory} 

\bibitem{Husimi1S} Crespi, B., Perez, G.  \& Chang, S-J.
\newblock Quantum Poincar\'e sections for two-dimensional billiards.
\newblock {\em Phys. Rev. E}, {\bf 47}, 986--991 (1993).

\bibitem{Husimi2S} Hentschel, M., Schomerus, H.  \& Schubert, R.
\newblock Husimi functions at dielectric interfaces: Inside-outside
duality for optical systems and beyond.
\newblock {\em Europhys. Lett.}, {\bf 62(5)}, 636--642 (2003).

\end{widetext}

\end{document}